\documentclass[%
 aip,
 amsmath,amssymb,
 reprint,%
]{revtex4-1}

\usepackage{graphicx}% Include figure files
\usepackage{dcolumn}% Align table columns on decimal point
\usepackage{bm}% bold math
\usepackage[utf8]{inputenc}
\usepackage[T1]{fontenc}
\usepackage{mathptmx}
\usepackage{etoolbox}
\usepackage{subcaption}
\usepackage{color}
\usepackage{CJKutf8}

\makeatletter
\def\@email#1#2{%
 \endgroup
 \patchcmd{\titleblock@produce}
  {\frontmatter@RRAPformat}
  {\frontmatter@RRAPformat{\produce@RRAP{*#1\href{mailto:#2}{#2}}}\frontmatter@RRAPformat}
  {}{}
}%
\makeatother
\begin{document}

\preprint{AIP/123-QED}
\begin{CJK*}{UTF8}{gbsn}
\title{Attention-Enhanced Neural Network Models for Turbulence Simulation}
% Force line breaks with \\
\author{Wenhui Peng (彭文辉)}
\affiliation{Department of Mechanics and Aerospace Engineering, Southern University of Science and Technology, Shenzhen 518055, China}
\affiliation{Southern Marine Science and Engineering Guangdong Laboratory (Guangzhou), Guangzhou 511458, China}
\affiliation{Guangdong-Hong Kong-Macao Joint Laboratory for Data-Driven Fluid Mechanics and Engineering Applications, Southern University of Science and Technology, Shenzhen 518055, China}
\affiliation{Department of Computer Engineering, Polytechnique Montreal, H3T1J4, Canada}

\author{Zelong Yuan (袁泽龙)}
\affiliation{Department of Mechanics and Aerospace Engineering, Southern University of Science and Technology, Shenzhen 518055, China}
\affiliation{Southern Marine Science and Engineering Guangdong Laboratory (Guangzhou), Guangzhou 511458, China}
\affiliation{Guangdong-Hong Kong-Macao Joint Laboratory for Data-Driven Fluid Mechanics and Engineering Applications, Southern University of Science and Technology, Shenzhen 518055, China}

\author{Jianchun Wang (王建春)}
\email{wangjc@sustech.edu.cn}
\affiliation{Department of Mechanics and Aerospace Engineering, Southern University of Science and Technology, Shenzhen 518055, China}
\affiliation{Southern Marine Science and Engineering Guangdong Laboratory (Guangzhou), Guangzhou 511458, China}
\affiliation{Guangdong-Hong Kong-Macao Joint Laboratory for Data-Driven Fluid Mechanics and Engineering Applications, Southern University of Science and Technology, Shenzhen 518055, China}

\date{\today}% It is always \today, today,

\begin{abstract}
Deep neural network models have shown a great potential in accelerating the simulation of fluid dynamic systems. Once trained, these models can make inference within seconds, thus can be extremely efficient. However, it becomes more difficult for neural networks to make accurate predictions when the flow becomes more chaotic and turbulent at higher Reynolds numbers. One of the most important reasons is that, existing models lack the mechanism to handle the unique characteristic of high-Reynolds-number turbulent flow: multi-scale flow structures are non-uniformly distributed and strongly nonequilibrium. In this work, we address this issue with the concept of visual attention: intuitively, we expect the attention module to capture the nonequilibrium of turbulence by automatically adjusting weights on different regions. We compare the model performance against a state-of-the-art neural network model as baseline, the Fourier Neural Operator (FNO), on two-dimensional (2D) turbulence prediction task. Numerical experiments show that the attention-enhanced neural network model outperforms existing state-of-the-art baselines, and can accurately reconstruct a variety of statistics and instantaneous spatial structures of turbulence at high Reynolds numbers flow. Furthermore, the attention mechanism provides 40\% error reduction with 1\% increase of parameters, at the same level of computational cost. 

\end{abstract}

\maketitle
\end{CJK*}

\section{\label{intro}Introduction}

Over the past few years, data-driven approaches based on machine learning (ML) have been extensively explored to complement and accelerate traditional computational fluid dynamics methods \cite{brunton2020machine,duraisamy2019turbulence}. Most of them follows two routes: ML-assisted model-driven approaches and pure data-driven approaches. 
The ML-assisted model-driven approaches aim to fit the closures of classical turbulence models based on consistency with high-resolution direct numerical simulations. Ling et al. presented a method of using deep neural networks to learn a model for the Reynolds stress anisotropy tensor from high fidelity simulation data \citep{ling2016reynolds}. Maulik et al. used neural networks to learn the map between inputs given by stencils of the vorticity and the streamfunction along with information from the eddy-viscosity kernels, and predict the sub-grid vorticity forcing in a temporally and spatially dynamic fashion \citep{maulik2019subgrid}. 
Wang et al. introduced a semi-explicit deep learning based framework for the reconstruction of the subgrid-scale stress in the large eddy simulation of incompressible turbulence \citep{wang2021artificial}. Beck et al. designed networks based on local convolution filters to predict the underlying unknown non-linear mapping from the coarse grid quantities to the closure terms without prior assumptions \citep{beck2019deep}. Portwood et al. showed that artificial neural networks can provide useful insights in the process of developing and improving turbulence closures \cite{portwood2021interpreting}. Park et al. used a fully connected neural network to develop a subgrid-scale (SGS) model in a turbulent channel flow \cite{park2021toward}. Yuan et al. developed a dynamic iterative approximate deconvolution models with Galilean invariance for the subgrid-scale stress in the
large-eddy simulation of turbulence \cite{yuan2021dynamic}.\\\indent Despite that the ML-assisted models are potentially more accurate than traditional turbulence models, they failed to achieve the desired computational expense reduction \cite{kochkov2021machine}. By contrast, the pure data-driven methods aim to approximate the entire Navier-Stokes equations by deep neural networks \cite{lusch2018deep,sirignano2018dgm,tang2021exploratory,sun2020neupde}. Once trained, the "black-box" neural network models can make inference within seconds on modern computers, thus can be extremely efficient compared with traditional approaches \cite{li2020neural}. \\ \indent Jin et al. proposed a fusion convolutional neural networks using measurements of the pressure field on the cylinder to predict the velocity field around a cylinder \cite{jin2018prediction}. Sekar et al. presented a data driven approach that combines the deep convolution neural network and deep multilayer perceptron to predict the incompressible laminar steady flow field over airfoils \cite{sekar2019fast}. Chen et al. proposed to apply the U-net architectures for fast prediction in laminar flows\cite{chen2019u}. Cheng et al. used the deep residual neural network to predict the flow velocity fields \cite{cheng2021deep}. Yousif et al. applied the generative adversarial network to reconstruct high-resolution turbulent flow fields \cite{yousif2021high}. Chen et al. proposed a graph neural networks architecture as a surrogate model for laminar flow prediction around two-dimensional obstacles \cite{chen2021graph}. Schmidt et al. simulated the turbulence quantities with a deep neural network to accelerate the modeling of transient flashing jets \cite{schmidt2021machine}. Han et al. designed a hybrid deep neural network architecture to capture the spatial-temporal features of unsteady flows from the high-dimensional numerical unsteady flow field \cite{han2019novel}. Nakamura et al. presented  a machine learning based reduced order model for three-dimensional complex flows, by combining a three-dimensional convolution neural network autoencoder and a long short-term memory network. Wang et al. developed a convolution neural network to improve the near-wall velocity field prediction \cite{wang2020predicting}. Li et al. proposed a deep neural network architecture for reconstructing the slices of the two-dimensional pressure field in three-dimensional flow, where the architecture embeds the convolution into the long short-term memory network to realize the purpose of using the upstream pressure to reconstruct downstream pressure \cite{li2021efficient}. Ren et al. proposed a deep learning framework that combines the long short-term memory network and the  convolution neural network for prediction of temporal evolution of turbulent flames \cite{ren2021predictive}. \\ \indent Despite these significant achievements in the acceleration of flow simulation, these models suffer from the generalization problem, and are sensitive to the parameters change\cite{kochkov2021machine}. Recent progress has been made to relieve the generalization problem with prior physical knowledge \cite{raissi2019physics}. Erichson et al. proposed to include the physics-informed prior knowledge to improve the model generalization performance on a fluid flow prediction task \cite{erichson2019physics}. Pawar et al. developed a physics-guided machine learning framework that leverages the interpretable physics-based model with a deep learning model, to improve the generalizability of neural networks \cite{pawar2021model}. Xu et al. employed the physics-informed deep learning by treating the governing equations as a parameterized constraint to reconstruct the missing flow dynamics\cite{xu2021explore}. Wang et al. further applied the physical constraints into the design of neural network, and proposed a grounded in principled physics model: the turbulent-flow network (TF-Net). The architecture of TF-Net contains trainable spectral filters in a coupled model of Reynolds-averaged Navier-Stokes simulation and large eddy simulation, followed by a specialized U-net for prediction. The TF-Net offers the flexibility of learned representations, and achieved state-of-the-art prediction accuracy \cite{wang2020towards}.\\\indent Proposed by Li et al., the Fourier neural operator (FNO) mimics the pseudo-spectral methods \cite{fan2019bcr,kashinath2020enforcing}: it parameterizes the integral kernel in the Fourier space, thus directly learns the mapping from any functional parametric dependence to the solution \cite{li2020fourier}. Benefited from the expressive and efficient architecture, the FNO outperforms previous state-of-the-art neural network models: the FNO achieves 1\% error rate on two-dimensional (2D) turbulence prediction task at low Reynolds numbers. However, when the fluid dynamic system becomes more chaotic, the Fourier neural operator suffers from the same problem \cite{li2020fourier}: the error rate arises over 15\% at the Reynolds number of $10^{5}$. 
\\\indent Making accurate predictions of turbulence at high Reynolds numbers has always been a difficult and challenging task for neural networks. An import reason is that the existing neural network models fail to efficiently incorporate the multi-scale and nonequilibrium characteristics of the high Reynolds numbers turbulence into the design of network architecture.

We propose to model the multi-scale and chaotic properties of turbulence with the concept of visual attention. Human visual attention allows us to focus on a specific region that contains important features while perceiving the surrounding environment with less concentration. Inspired by the human visual attention, we expect the attention mechanism to capture strongly nonequilibrium regions by focusing on certain important regions in the flow field. 

The attention mechanism in deep learning was first introduced by Bahdanau et al. for machine translation \cite{bahdanau2014neural}. In recent years, attention mechanism has shown itself to be very successful in boosting the performance of neural networks on a variety of tasks, ranging from nature language processing to computer vision \cite{vaswani2017attention, parmar2018image,liu2018visual}. In fluid dynamics, attention mechanism has also been used to enhance the reduced order model to extract temporal feature relationships from high-fidelity numerical solutions \cite{wu2021reduced}. In this work, we couple the attention mechanism with the state-of-the-art FNO model to improve the model performance on high Reynolds numbers flow.

This paper is organized as follow: section \ref{fno_intro} briefly introduces the Fourier neural operator, section \ref{attention_intro} shows the detailed implementation of the attention mechanism, in section \ref{attention_exp} we benchmark the attention improvement with numerical experiments on the prediction task of 2D turbulence. In the sections \ref{discussion} and \ref{conclusion} we give discussions and draw conclusions, respectively.

\section{The Fourier neural operator}\label{fno_intro}
Most neural network architectures have focused on learning mappings between finite-dimensional Euclidean spaces, they are good at learning a single instance of the equation, but have difficulty to generalize well once the governing equation parameters or boundary conditions changes \cite{raissi2019physics,pan2020physics,wu2020data,xu2021deep}.
Given a finite collection of the observed input-output pairs, the Fourier neural operator learns the mapping from any functional parametric dependence to the solution, meaning that they learn an entire family of partial differential equations instead of a single equation \cite{li2020fourier}. Specifically, let $D \subset \mathbb{R}^{d}$ be a bounded, open set, and notate the target non-linear mapping as $G^{\dagger}: \mathcal{A}\rightarrow \mathcal{U}$, where $\mathcal{A}\left(D ; \mathbb{R}^{d_{a}}\right)$ and $\mathcal{U}\left(D ; \mathbb{R}^{d_{u}}\right)$ are separable Banach spaces of function taking values in $\mathbb{R}^{d_{a}}$ and $\mathbb{R}^{d_{u}}$ respectively \cite{beauzamy2011introduction}. The Fourier neural operators learns an approximation of $G^{\dagger}$ by constructing a mapping parameterized by $\theta \in \Theta$.

$$
G: \mathcal{A} \times \Theta \rightarrow \mathcal{U}.
$$

The optimal parameters $\theta^{\dagger} \in \Theta$ are determined in the test-train setting by using a data-driven empirical approximation \cite{vapnik1999overview}, such that $G\left(\cdot, \theta^{\dagger}\right)=G_{\theta^{\dagger}} \approx G^{\dagger}$.

\begin{figure*}
\centering
\includegraphics[width=.9\linewidth]{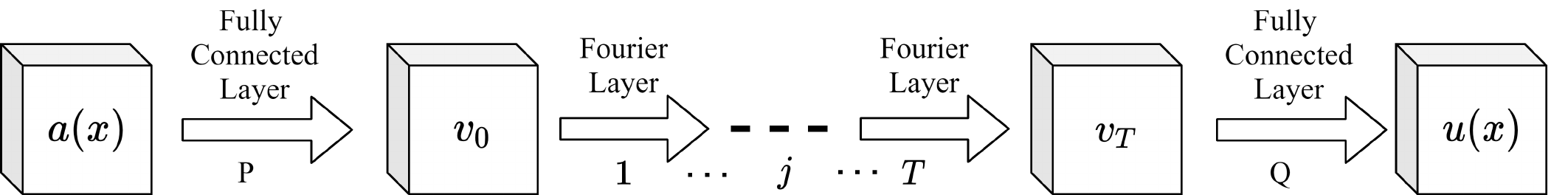}
\caption{Fourier neural operator (FNO) architecture.}
\label{fno_architechture}
\end{figure*}

The neural operators \cite{li2020neural} are formulated as an iterative architecture $v_{0} \mapsto$ $v_{1} \mapsto \ldots \mapsto v_{T}$ where $v_{j}$ for $j=0,1, \ldots, T-1$ is a sequence of functions each taking values in $\mathbb{R}^{d_{v}}$, as shown in Fig.\ref{fno_architechture}.  Firstly, the input $a \in \mathcal{A}$ is transformed to a higher dimensional representation 
 $$
 v_{0}(x)=P(a(x)).
 $$
 by a fully connected layer $P$, then the higher dimensional representation is updated iteratively by Eq.(\ref{eq:update}), where $\mathcal{K}:\mathcal{A} \times \Theta_{\mathcal{K}} \rightarrow \mathcal{L}\left(\mathcal{U}\left(D ; \mathbb{R}^{d_{v}}\right), \mathcal{U}\left(D ; \mathbb{R}^{d_{v}}\right)\right)$ maps to bounded linear operators on $\mathcal{U}\left(D ; \mathbb{R}^{d_{v}}\right)$ and is parameterized by $\phi \in \Theta_{\mathcal{K}}$, $W: \mathbb{R}^{d_{v}} \rightarrow \mathbb{R}^{d_{v}}$ is a linear transformation, and $\sigma: \mathbb{R} \rightarrow \mathbb{R}$ is an elementally defined non-linear activation function. Lastly, the output $u \in \mathcal{U}$ is obtained by applying the local transformation $u(x)=Q\left(v_{T}(x)\right)$, where $Q: \mathbb{R}^{d_{v}} \rightarrow \mathbb{R}^{d_{u}}$.
 
\begin{equation}\label{eq:update}
v_{t+1}(x) =\sigma\left(W v_{t}(x)+\left(\mathcal{K}(a ; \phi) v_{t}\right)(x)\right), \quad \forall x \in D.
\end{equation}
 
Let $\mathcal{F}$ and $\mathcal{F}^{-1}$ denote the Fourier transform and its inverse transform of a function $f: D \rightarrow \mathbb{R}^{d_{v}}$ respectively. Replacing the kernel integral operator in Eq. (\ref{eq:update}) by a convolution operator defined in Fourier space, and applying the convolution theorem, the Fourier integral operator can be expressed by Eq.(\ref{eq:k}), where $R_{\phi}$ is the Fourier transform of a periodic function $\kappa: \bar{D} \rightarrow \mathbb{R}^{d_{v} \times d_{v}}$ parameterized by $\phi \in \Theta_{\mathcal{K}}$, as shown in Fig.\ref{fno_architechture}. 

\begin{equation}\label{eq:k}
\left(\mathcal{K}(\phi) v_{t}\right)(x)=\mathcal{F}^{-1}\left(R_{\phi} \cdot\left(\mathcal{F} v_{t}\right)\right)(x), \quad \forall x \in D.
\end{equation}

The frequency mode $k \in D$ is assumed to be
periodic, and it allows a Fourier series expansion, which expresses as the discrete modes $k \in \mathbb{Z}^{d}$. The finite-dimensional parameterization is implemented by truncating the Fourier series at a maximal number of modes $k_{\max }=\left|Z_{k_{\max }}\right|=\mid\left\{k \in \mathbb{Z}^{d}:\left|k_{j}\right| \leq k_{\max , j}\right.$, for $\left.j=1, \ldots, d\right\} \mid$. We discretize the domain $D$ with  $n \in \mathbb{N}$ points, where $v_{t} \in \mathbb{R}^{n \times d_{v}}$ and $\mathcal{F}\left(v_{t}\right) \in \mathbb{C}^{n \times d_{v}}$. $R_{\phi}$ is parameterized as complex-valued weight tensor containing a collection of truncated Fourier modes $R_{\phi} \in \mathbb{C}^{k_{\max } \times d_{v} \times d_{v}}$, and $\mathcal{F}\left(v_{t}\right) \in \mathbb{C}^{k_{\max } \times d_{v}}$ is obtained by truncating the higher modes, therefore

\begin{eqnarray*}
&&\left(R_{\phi}\cdot\left(\mathcal{F} v_{t}\right)\right)_{k, l}=\sum_{j=1}^{d_{v}} R_{\phi  k, l, j}\left(\mathcal{F} v_{t}\right)_{k, j}, \\&&\quad k=1, \ldots, k_{\max }, \quad j=1, \ldots, d_{v}.
\end{eqnarray*}

In CFD modeling, the flow is typically uniformly discretized with resolution $s_{1} \times \cdots \times s_{d}=n$, and $\mathcal{F}$ can be replaced by the Fast Fourier Transform (FFT). For $f \in \mathbb{R}^{n \times d_{v}}, k=\left(k_{1}, \ldots, k_{d}\right) \in \mathbb{Z}_{s_{1}} \times \cdots \times \mathbb{Z}_{s_{d}}$, and $x=\left(x_{1}, \ldots, x_{d}\right) \in D$ the FFT $\hat{\mathcal{F}}$ and its inverse $\hat{\mathcal{F}}^{-1}$ are given by Eq. (\ref{eq:fft}), for $l=1, \ldots, d_{v}$.

\begin{equation}\label{eq:fft}
\begin{aligned}
&(\hat{\mathcal{F}} f)_{l}(k)=\sum_{x_{1}=0}^{s_{1}-1} \cdots \sum_{x_{d}=0}^{s_{d}-1} f_{l}\left(x_{1}, \ldots, x_{d}\right) e^{-2 i \pi \sum_{j=1}^{d} \frac{x_{j} k_{j}}{s_{j}}}. \\
&\left(\hat{\mathcal{F}}^{-1} f\right)_{l}(x)=\sum_{k_{1}=0}^{s_{1}-1} \cdots \sum_{k_{d}=0}^{s_{d}-1} f_{l}\left(k_{1}, \ldots, k_{d}\right) e^{2 i \pi \sum_{j=1}^{d} \frac{x_{j} k_{j}}{s_{j}}}.
\end{aligned}
\end{equation}

\section{Attention-enhanced neural network}\label{attention_intro}

Human visual attention allows us to focus on a specific region that contains important features while perceiving the surrounding environment with less concentration. Inspired by the concept of visual attention, we propose to model the nonequilibrium features of turbulence with the self-attention module in machine vision\cite{zhang2019self}. 

\begin{figure*}
\centering
\includegraphics[width=0.9\textwidth]{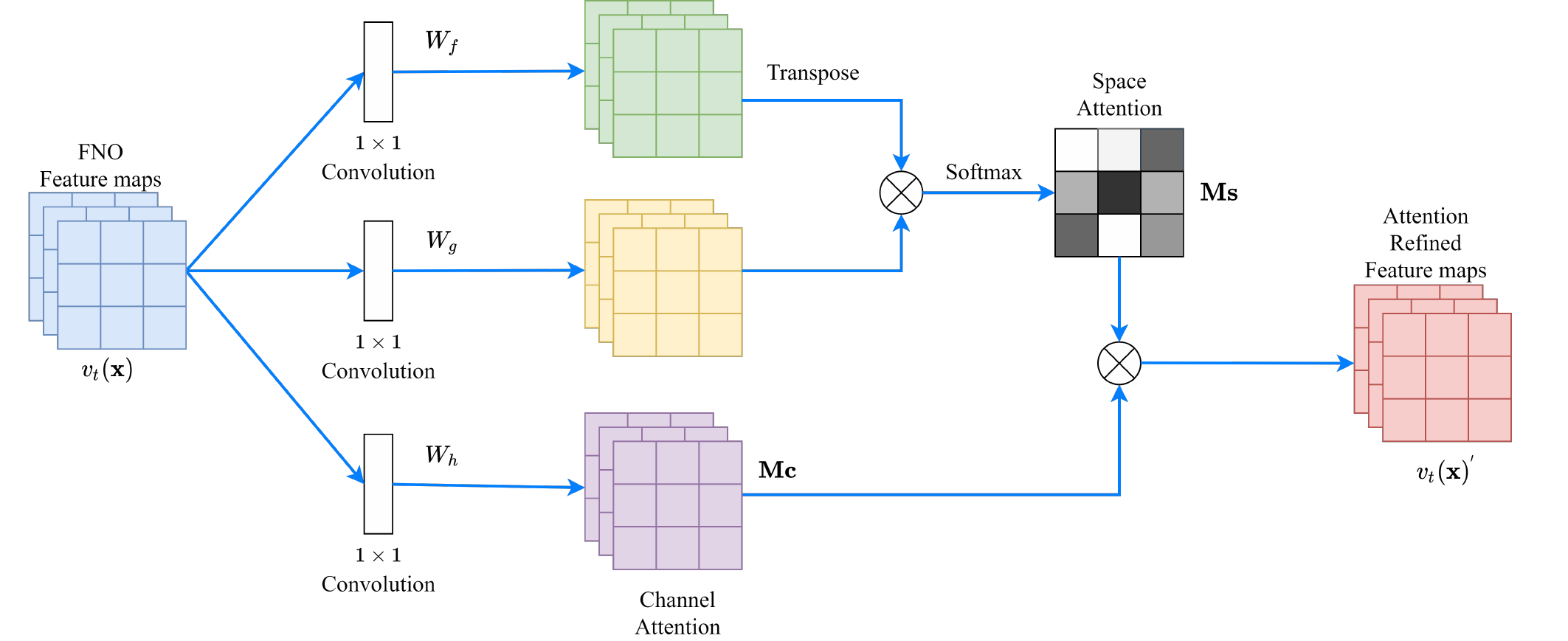}
\caption{Self-attention block.}
\label{attention_block}
\end{figure*}

The idea of attention mechanism was first popularly applied in the domain of natural language processing \cite{vaswani2017attention}. An attention function can be described as mapping a query and a set of key-value pairs to an output, where the query, keys, values, and output are all vectors. The output is computed as a weighted sum of the values, where the weight assigned to each value is computed by a compatibility function of the query with the corresponding key \cite{vaswani2017attention}. The three sub-modules (query, key, and value) are the pivotal components of attention mechanism, which come from the concepts of information retrieval systems \cite{kowalski2007information}. An example is the query search, where the search engine maps the query to a set of keys, and output the values. Zhang et al. extended the idea into the domain of computer vision by proposing the self-attention module \cite{zhang2019self}. The overall architecture of the self-attention block is shown in Fig.\ref{attention_block}, it learns sequential attention maps that are called the channel attention and the space attention. The first layer performs $1\times1$ convolutions on $v_{t} (\boldsymbol{x})$ to sequentially infer the channel attention map $\mathbf{Mc}$ and the space attention map $\mathbf{Ms}$, as shown in Eq.\ref{eq3}. 
The self-attention operation mimics the "proportional retrieval" approach from information retrieval systems, where $s_{i j}$ corresponds to the probability vector \cite{kowalski2007information}.

\begin{eqnarray}
\mathbf{Mc} &=& W_{h} v_{t}(\boldsymbol{x}), \nonumber\\
\mathbf{Ms}&=&\frac{\exp \left(s_{i j}\right)}{\sum_{i=1}^{N} \exp \left(s_{i j}\right)}, \nonumber\\
\text { where } s_{i j}&=&\left(W_{f} v_{t}(\boldsymbol{x}_{\boldsymbol{i}})\right)^{T} \left(W_{g} v_{t}(\boldsymbol{x}_{\boldsymbol{j}})\right).
\label{eq3}
\end{eqnarray}

The convolution parameters $W_{f},W_{g},W_{h}$ learn the embedding of query, key and value, respectively. Eq. \ref{eq4} gives the attention refined feature maps, where $\otimes$ denotes the dot product.
\begin{equation}
v_{t}(\boldsymbol{x})^{'}=\mathbf{M}_{\mathbf{s}} \otimes \mathbf{Mc}.
\label{eq4}
\end{equation}

\begin{figure*}
\centering
\includegraphics[width=.65\textwidth]{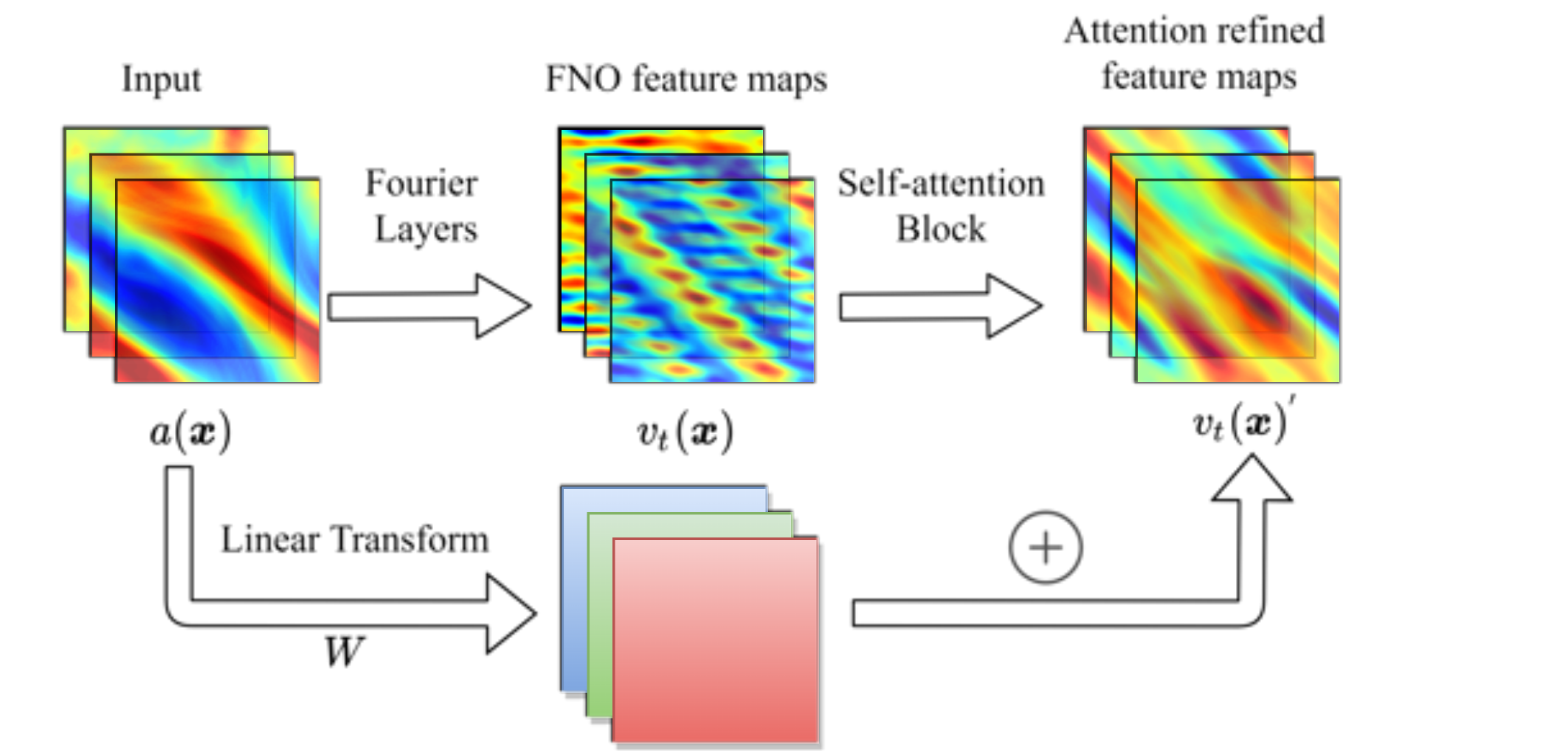}
\caption{Attention enhanced Fourier neural operator (FNO).}
\label{fno_attention}
\end{figure*}

The sequence of functions $v_{0} \mapsto$ $v_{1} \mapsto \ldots \mapsto v_{T}$ is the core structure of Fourier network, which directly affects the model learning capability. In the original network, $v_{t} \mapsto v_{t+1}$ is updated by Eq. \ref{eq:update}, and we use
the self-attention block to refine the FNO output as feature augmentation as shown in Fig.\ref{fno_attention}. The self-attention block takes the feature maps of Fourier neural operators $v_{t} (\boldsymbol{x})$ as input, and output the attention refined feature maps $v_{t} (\boldsymbol{x})^{'}$. 

Despite that the self-attention module allows feature refinement for any neural operators $v_{t} \Rightarrow v_{t}^{'}$ for $t=1, \ldots, T$, our numerical experiments show that a single block at the end of sequence $v_{T} \Rightarrow v_{T}^{'}$ has the same level of improvement as refining all neural operators in the sequence $v_{1} \Rightarrow$ $v_{1}^{'} \mapsto \ldots \mapsto v_{T}\Rightarrow v_{T}^{'}$.

The attention parameters $W_{f},W_{g},W_{h}$ can be jointly learned with the Fourier layer during training. More importantly, since the attention block adopts $1\times1$ convolution filters, it retains the mesh invariant property of neural operators, meaning that the attention-enhanced FNO still keeps the ability of training on coarse grid and making inference on finer grid without seeing any higher resolution data. Details are 
discussed in Sec. \ref{mesh}.

\section{Performance Benchmark}\label{attention_exp}

In this section, we benchmark the performance of attention-augmented neural network against the original FNO model with various criteria, including the vorticity prediction error, the vorticity spectrum, the probability density function (PDF) for vorticity and the condition average error of the vorticity. We conduct numerical experiments to evaluate the prediction ability of the two models (FNO, FNO+Attention) on the same dataset of 2D incompressible turbulence with reference \cite{li2020fourier}. The numerical experiments are implemented on the Pytorch and MindSpore deep learning frameworks. We also compare the performance of two models on more complex turbulent flows: the Taylor-Green turbulence and free shear turbulence. Specifically, we are interested in comparing the prediction accuracy of the two models at different time steps and at different Reynolds numbers.

% The experiments are ran on the MindSpore deep learning framework. 

\subsection{Dataset description}

Eq. (\ref{eq:ns})
describes the dimensionless Navier-Stokes equations for a viscous, incompressible fluid on the unit torus, where $u \in C\left([0, T] ; H_{\text {per }}^{r}\left((0,1)^{2} ; \mathbb{R}^{2}\right)\right)$ for any $r>0$ is the velocity field, $\omega=\nabla \times \mathbf{u}$ is the vorticity, $\omega_{0} \in L_{\mathrm{per}}^{2}\left((0,1)^{2} ; \mathbb{R}\right)$ is the initial vorticity, $Re$ is the Reynolds number, and $f \in$ $L_{\mathrm{per}}^{2}\left((0,1)^{2} ; \mathbb{R}\right)$ is the forcing function.

\begin{equation}\label{eq:ns}
\begin{aligned}
\partial_{t} \omega\left( {{\mathbf{x}},t} \right)+\mathbf{u}\left( {{\mathbf{x}},\normalfont{t}} \right) \cdot \nabla \omega(\mathbf{x}, t) &=(1/Re) \Delta \omega(\mathbf{x}, t)+f(\mathbf{x}),\\
% \nabla \cdot u(\mathbf{x}, t) &=0, \\
\omega\left( {{\mathbf{x}},t} \right) &=\omega_{0}(\mathbf{x}), \\ \mathbf{x} \in(0,1)^{2}, &~t \in[0, T].
\end{aligned}
\end{equation}

The initial condition $\omega_{0}(\mathbf{x})$ is generated according to $\omega_{0} \sim \mu$ with periodic boundary conditions\cite{li2020fourier}, where the Gaussian random distribution $\mu=\mathcal{N}\left(0,7^{3 / 2}(49-\Delta)^{-2.5}\right)$ is expressed as
\begin{equation}\label{normal}
\mu \left( {\mathbf{x}} \right) = \frac{1}{{2\pi  \times {7^3}{{\left( {49 - \Delta } \right)}^{ - 5}}}}\exp \left[ { - \frac{{x_1^2 + x_2^2}}{{2 \times {7^3}{{\left( {49 - \Delta } \right)}^{ - 5}}}}} \right].
\end{equation}
Here, $\Delta =2\pi/N$ represents the uniform grid spacing.

\begin{figure*}
\centering
\includegraphics[width=1\textwidth]{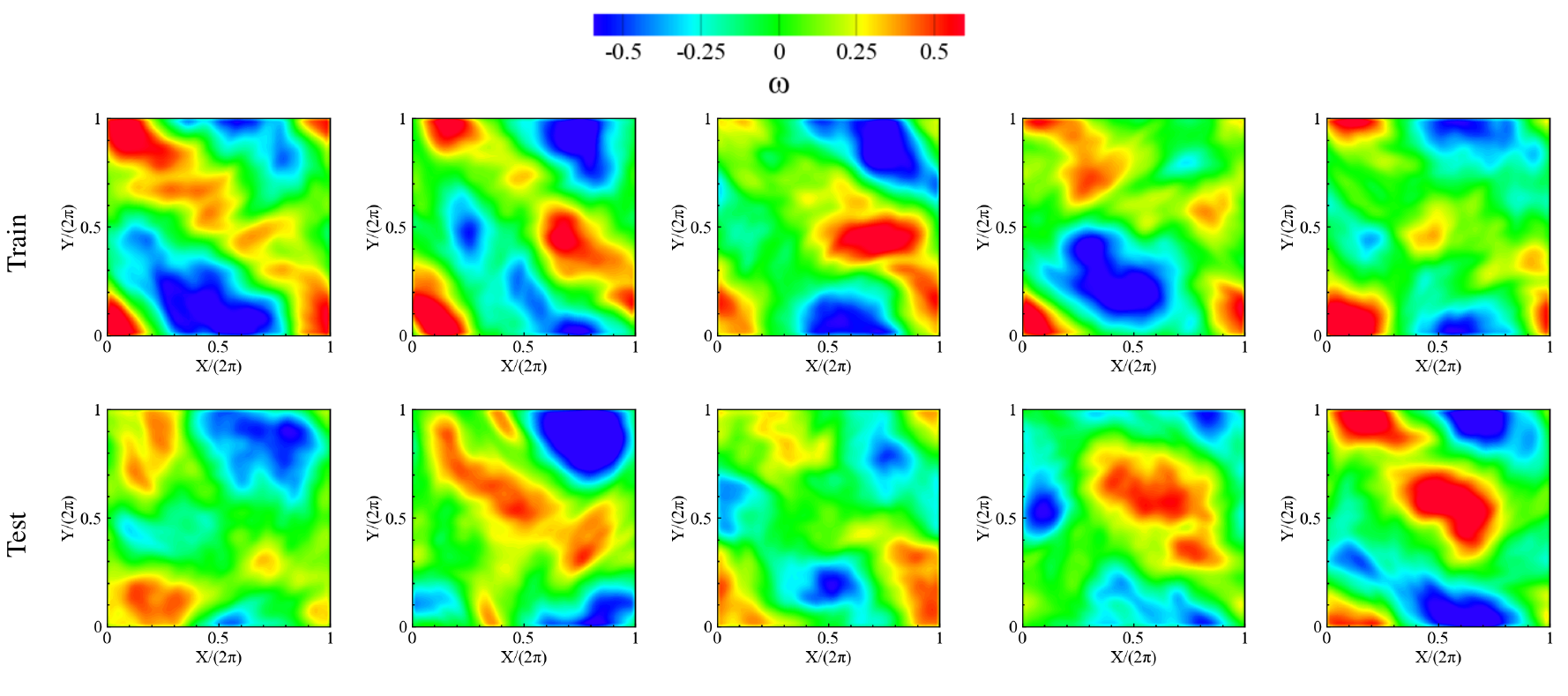}
\caption{Random sampled initial conditions from the training set and testing set.}
\label{initialization}
\end{figure*}

Fig. \ref{initialization} shows the initial vorticity of ten random conditions, which are  sampled from the training set and the testing set. The forcing is kept fixed $f(\mathbf{x})=0.1\left[\sin \left(2 \pi\left(x_{1}+x_{2}\right)\right)+\right.$ $\left.\cos \left(2 \pi\left(x_{1}+x_{2}\right)\right)\right]$.

In this paper, a pseudo-spectral method is applied to numerically simulate the incompressible 2D homogeneous isotropic turbulence in a cubic box of $(2\pi)^2$ on a uniform grid with periodic boundary conditions. The vorticity can be expanded as the Fourier series,
\begin{equation}\label{eq:w_hat}
\omega \left( {{\mathbf{x}},t} \right) = \sum\limits_{\mathbf{k}} {\hat \omega \left( {{\mathbf{k}},t} \right)\exp \left( {i{\mathbf{k}}\cdot{\mathbf{x}}} \right)} ,
\end{equation}
where $i$ stands for the imaginary unit, namely, $i^2=-1$, $\mathbf{k}=[k_x, k_y]^T$ is the wavenumber vector, and the superscript ``$\widehat  \bullet$" represents the variable in Fourier space. The vorticity equation in wavenumber space can be derived as
\begin{equation}\label{eq:ns_wave}
\left( {\frac{d}{{dt}} + \frac{1}{{{\mathop{\rm Re}\nolimits} }}{k^2}} \right)\hat \omega \left( {{\mathbf{k}},t} \right) = \hat f\left( {{\mathbf{k}},t} \right) - \widehat {{\mathbf{u}}\left( {\mathbf{x},t} \right) \bullet \nabla \omega \left( {\mathbf{x},t} \right)},
\end{equation}
The nonlinear advection term is calculated by the pseudo-spectral method \cite{li2020fourier}. The basic procedure is to first transform the Fourier variables $ {\mathbf{\hat u}}\left( {{\mathbf{k}},t} \right) = {\left[ { - i{k_y}\hat \omega /{k^2},\;\;i{k_x}\hat \omega /{k^2}} \right]^T}$ and $\hat \omega \left( {{\mathbf{k}},t} \right)$ to ${{\mathbf{u}}\left( {\mathbf{x},t} \right)}$ and ${\nabla \omega \left( {\mathbf{x},t} \right)}$ in physical space by the inverse fast Fourier transform, and perform the multiplication in physical space to obtain $ {{\mathbf{u}}\left( {\mathbf{x},t} \right) \bullet \nabla \omega \left( {\mathbf{x},t} \right)}$, after that the forward fast Fourier transform is calculated and aliasing error is removed by the 3/2 zero-padding rule\cite{li2020fourier}, respectively. Data are generated on the grid size of $256\times256$  and are downsampled to $64 \times 64$. Time is advanced with the Crank-Nicolson scheme, where the time-step is set to be $1 0^{-4}$, and the solution is recorded every $t=1$ time units. An explicit two-step Adams-Bashforth scheme is selected as the time marching scheme with second-order temporal accuracy. For a partial differential equation $\partial_t \hat \omega = R(\hat \omega, t)$, the iterative scheme for time advancement is given by Eq. \ref{eq:scheme_ab2}, where $\Delta t$ is the time step, $t_n = n \Delta t$, and $\hat \omega^n$ is the vorticity at time $t_n$.

\begin{equation}\label{eq:scheme_ab2}
	\begin{aligned}
	{\hat \omega ^{n + 1}} = {\hat \omega ^n} + \Delta t\left[ {\frac{3}{2}R\left( {{\hat \omega ^n},{t_n}} \right) - \frac{1}{2}R\left({{\hat \omega ^{n - 1}},{t_{n - 1}}} \right)} \right]
	\end{aligned}.
\end{equation}

\subsection{Performance benchmark at different time steps}
Since the model prediction errors are produced and accumulated at every step, prediction error increases dramatically with time due to the chaotic features of turbulence. Reducing the accumulated prediction errors on time dimension is therefore still one of the most challenging problems for surrogate models.

In this numerical experiment, we generate 1200 pairs of input-output data with the numerical solver, where each sample contains 20 steps of solutions of a random initialized condition at Reynolds number $10^5$. The solution is recorded every $t=1$ time units. Both models (FNO, FNO+Attention) take the vorticity at previous 10 time steps solutions as input, and gives the vorticity at the next time step as output. During training, the vorticity of first 10 steps $\omega|_{(0,1)^{2} \times[0,10]}$ is stacked over temporal dimension as the model input, and the model
recurrently predicts the vorticity at the next step to fit the vorticity at next 10 steps $\omega|_{(0,1)^{2} \times[11,20]}$, which are labeled as ground truth.

\begin{figure*}
\centering
\includegraphics[width=1\textwidth]{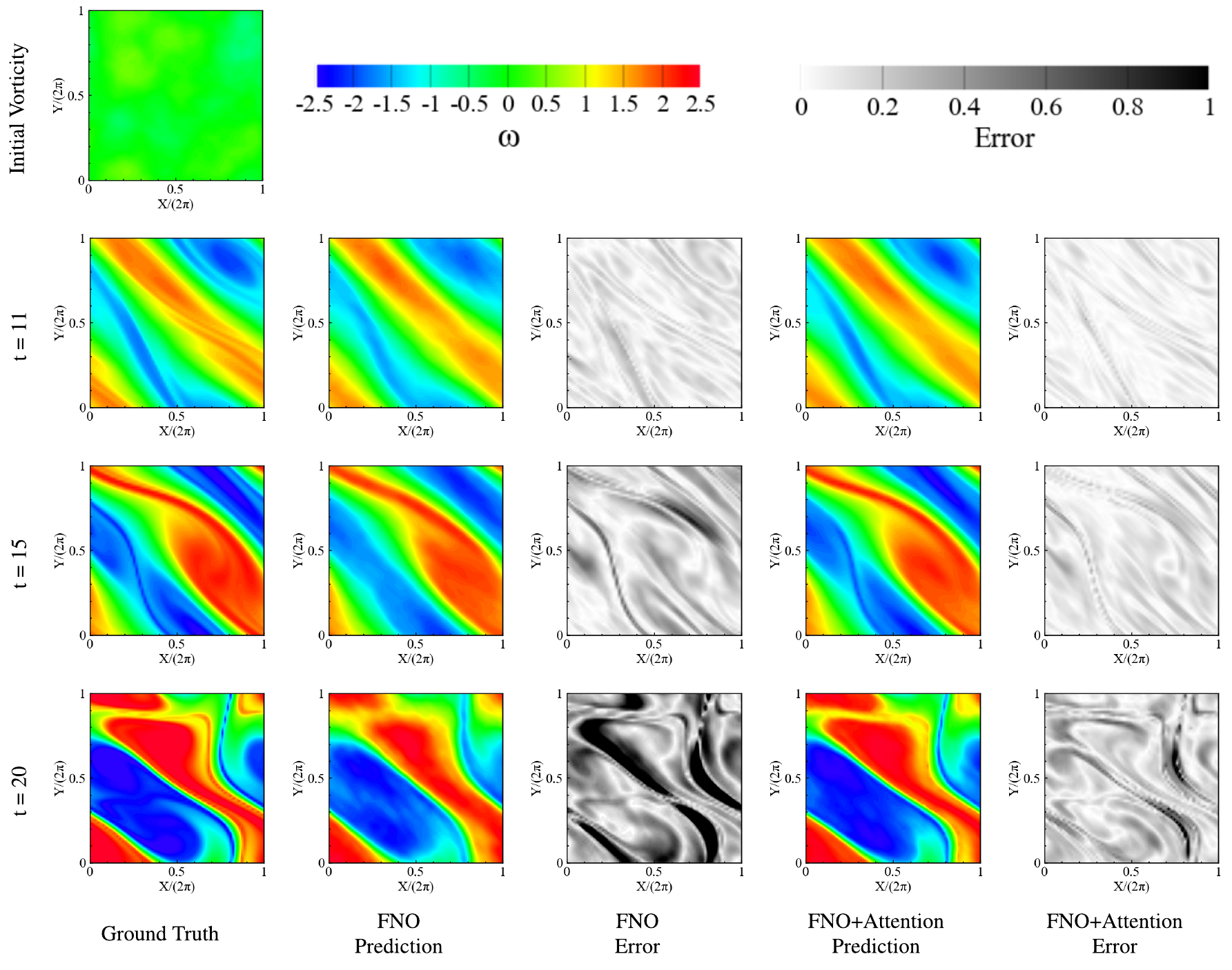}
\caption{Vorticity prediction and absolute error at selected time steps.}
\label{vor_time_steps}
\end{figure*}

\begin{figure*}
\centering
\includegraphics[width=1\textwidth]{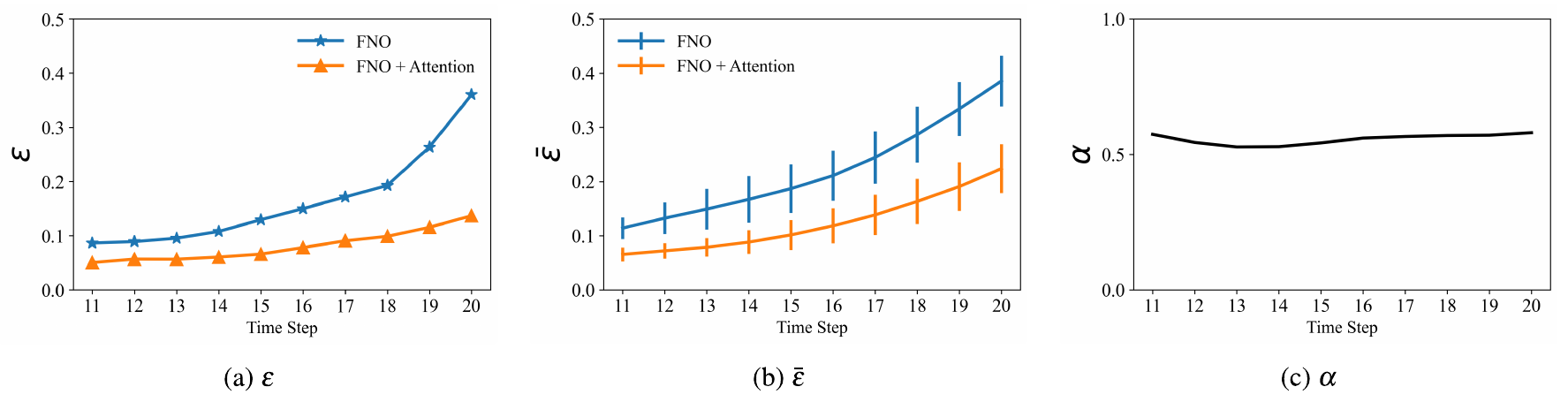}
\textbf{(a)} Spatial-averaged relative error of vorticity on single test sample. \textbf{(b)} Spatial-averaged relative error of vorticity on 200 test samples. \textbf{(c)} Ratio of mean relative error.
\caption{Relative error comparison at consequent time steps.}
\label{error_time_steps2}
\end{figure*}

It is worth noting that the predicted vorticity at each step is recurrently treated as ground truth and reused as the inputs with the advance of time, thus the prediction errors are accumulated iteratively. We use 1000 samples for training and 200 samples for testing. After training, we evaluate both models on the test dataset, and compare their performance at three selected time steps (t=11, t=15, t=20). Fig. \ref{vor_time_steps} compares the predicted vorticity and the absolute errors of a test sample: both models can accurately reconstruct the instantaneous spatial structures of turbulence in the beginning, however, the difference is enlarged significantly as time progresses. The FNO error increases at regions where the vorticity has dramatically changes. In contrast, the errors of attention-enhanced FNO are visibly smaller in terms of the region and magnitude. We find the same phenomenon in Fig.\ref{error_time_steps2} as we investigate the relative error $\epsilon$ of each time step. The 
relative error $\epsilon$ is defined by Eq. \ref{re_error}, where $\hat{\omega}$ is the predicted vorticty and $\omega$ is the ground truth vorticity. 

\begin{equation}\label{re_error}
\epsilon = \frac{\|\hat{\mathrm{\omega}}-\mathrm{\omega}\|_{2}}{\|\mathrm{\omega}\|_{2}}, 
\text { where }\|\mathrm{x}\|_{2}=1/n
{\sqrt{\sum_{k=1}^{n}\left|x_{k}\right|^{2}}}.
\end{equation}

\begin{figure*}
\centering
\includegraphics[width= 1\textwidth]{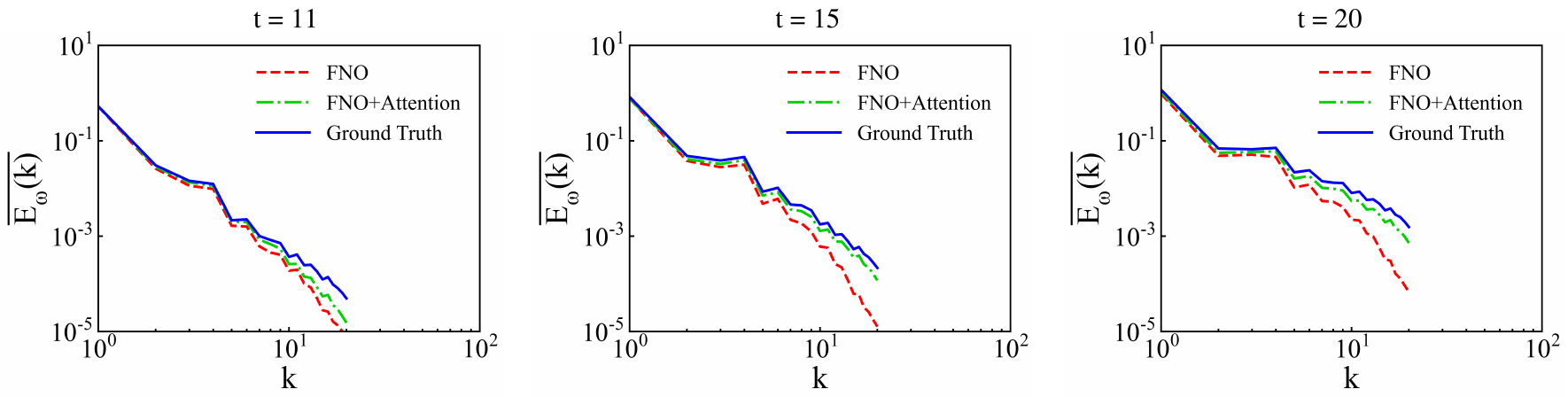}
\caption{Averaged vorticity spectrum on 200 test samples.}
\label{spectra_time_steps_avg}
\end{figure*}

\begin{figure*}
\centering
\includegraphics[width= 1\textwidth]{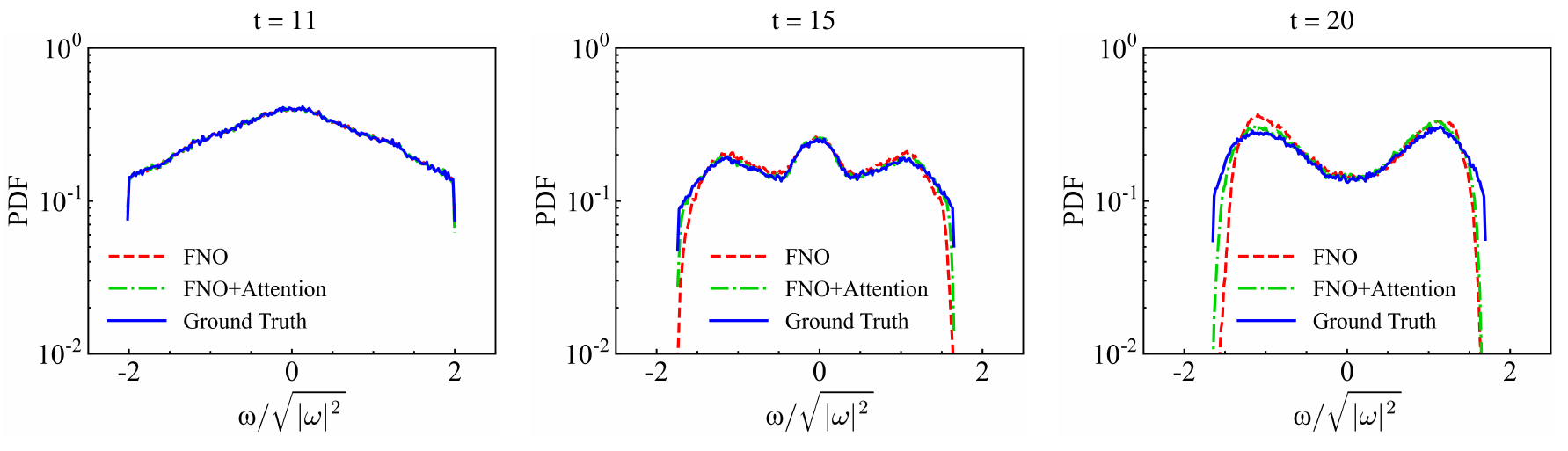}
\caption{PDF of normalized vorticity at different time steps.}
\label{pdf_time}
\end{figure*}

The ensemble-averaged vorticity spectrum using 200 test samples $\overline{E_{\omega}(k)}$ is also compared in Fig.\ref{spectra_time_steps_avg}: the predicted vorticity spectrum of both models can agree well with the ground truth in the  low-wave number region. However, the FNO predicted spectrum deviates significantly from the ground truth at high-wave number region as time advances. In contrast, the attention enhanced model can accurately capture the small-scale flow structures and well reconstruct the vorticity spectrum at different flow scales.
Fig.\ref{pdf_time} illustrates the PDFs of the normalized predicted vorticity on different time steps, where the vorticity is normalized by the root-mean-square values of ground truth vorticity. The predictions of both models have a good agreement with the ground truth in the beginning. However, as time advances, the predicted PDFs of both models become narrower, with the attention-enhanced FNO being closer to the ground truth. 

\begin{figure*}
\centering
\includegraphics[width=1\textwidth]{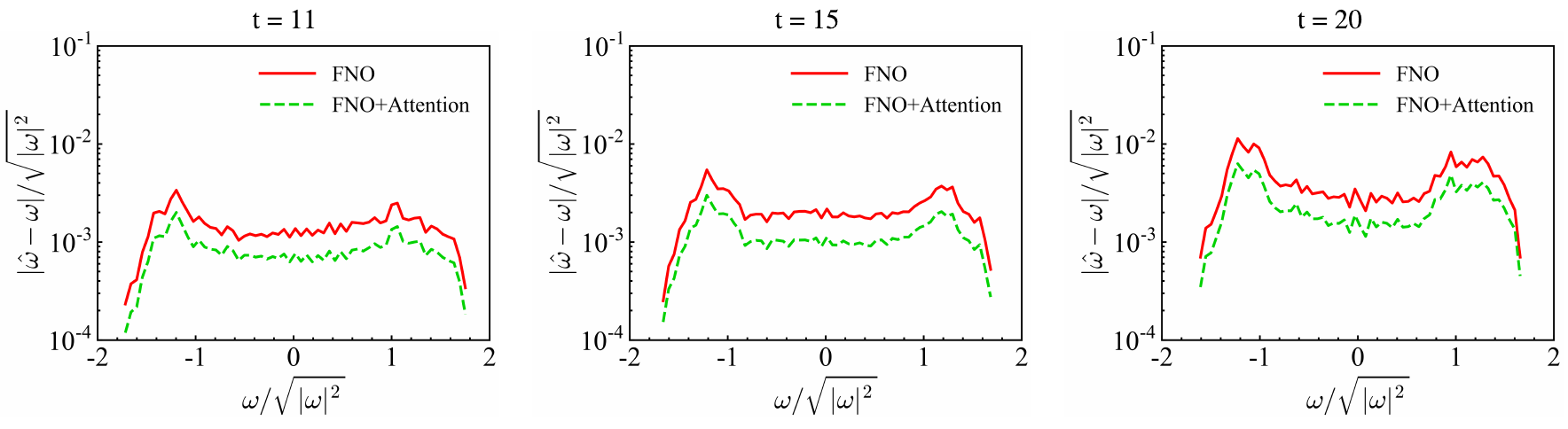}
\caption{Condition average error of vorticity at different time steps.}
\label{error_con_avg_time}
\end{figure*}

Fig.\ref{error_con_avg_time} shows the average of normalized absolute error conditioned on normalized vorticity at different time steps, where both vorticity and absolute error are normalized by the root-mean-square values of ground truth vorticity. We notice that the prediction errors of both models increase with the advance of time, and the attention-enhanced FNO achieves better performance with smaller error in the whole vorticity range.
As can be seen from Fig.\ref{error_time_steps2}, where the spatial-averaged relative errors are plotted, the attention-enhanced model has achieved over 40\% error reduction throughout all time steps.

\subsection{Performance benchmark at different Reynolds numbers}
Since the multi-scale flow structures become more chaotic and turbulent with the increasing of Reynolds numbers, reconstruction of turbulence at high Reynolds numbers has always been a challenging task for neural networks.\\\indent Table \ref{baseline_table} shows the 
prediction relative errors (defined in Eq.\ref{re_error}) of existing state-of-the-art neural network models including the FNO \cite{li2020fourier}. Note that the results of U-Net\cite{ronneberger2015u}, TF-Net\cite{wang2020towards} and Resnet \cite{he2016deep} are cited from reference\cite{li2020fourier}. It is noted that all existing models perform well at low Reynolds number ($Re= 10^3$), and fails at higher Reynolds numbers ($Re= 10^4$ and $Re= 10^5$), with FNO achieving the lowest error. In contrast, the attention-enhanced FNO model further reduces the prediction errors of FNO by 40\% at higher Reynolds numbers.

\begin{table*}
\caption{Prediction errors of neural network models at different Reynolds numbers}
\label{baseline_table}
\begin{ruledtabular}
\begin{tabular}{l|r|cccc}
Model & Parameters & $Re = 10^3$ & $Re = 10^4$ & $Re = 10^5$ \\
\hline FNO + Attention & $466,222$ & $0.0096$ & $0.0855$ & $0.0864$ \\
FNO\cite{li2020fourier} & 465,717  & $0.0094$ & $0.1479$ & $0.1507$ \\
% U-Net + Attention & $25,399,597$ & $0.0245$ & $0.2051$ & $0.1982$ \\
U-Net\cite{chen2019u} & $24,950,491$ & $0.0245$ & $0.2051$ & $0.1982$ \\
TF-Net\cite{wang2020towards}& $7,451,724$ & $0.0225$ & $0.2253$  & $0.2268$ \\
ResNet\cite{he2016deep} & 266,641 & $0.0701$ & $0.2871$ & $0.2753$ \\
\hline
\end{tabular}
\end{ruledtabular}
\end{table*}

Here, we benchmark the prediction performance of FNO and attention-enhanced FNO, at the same time step (t=15), for different Reynolds numbers. We train and test both models on three groups of data where the flow Reynolds number is set to $10^{3},10^{4},10^{5}$ respectively. 

\begin{figure*}
\centering
\includegraphics[width=1\textwidth]{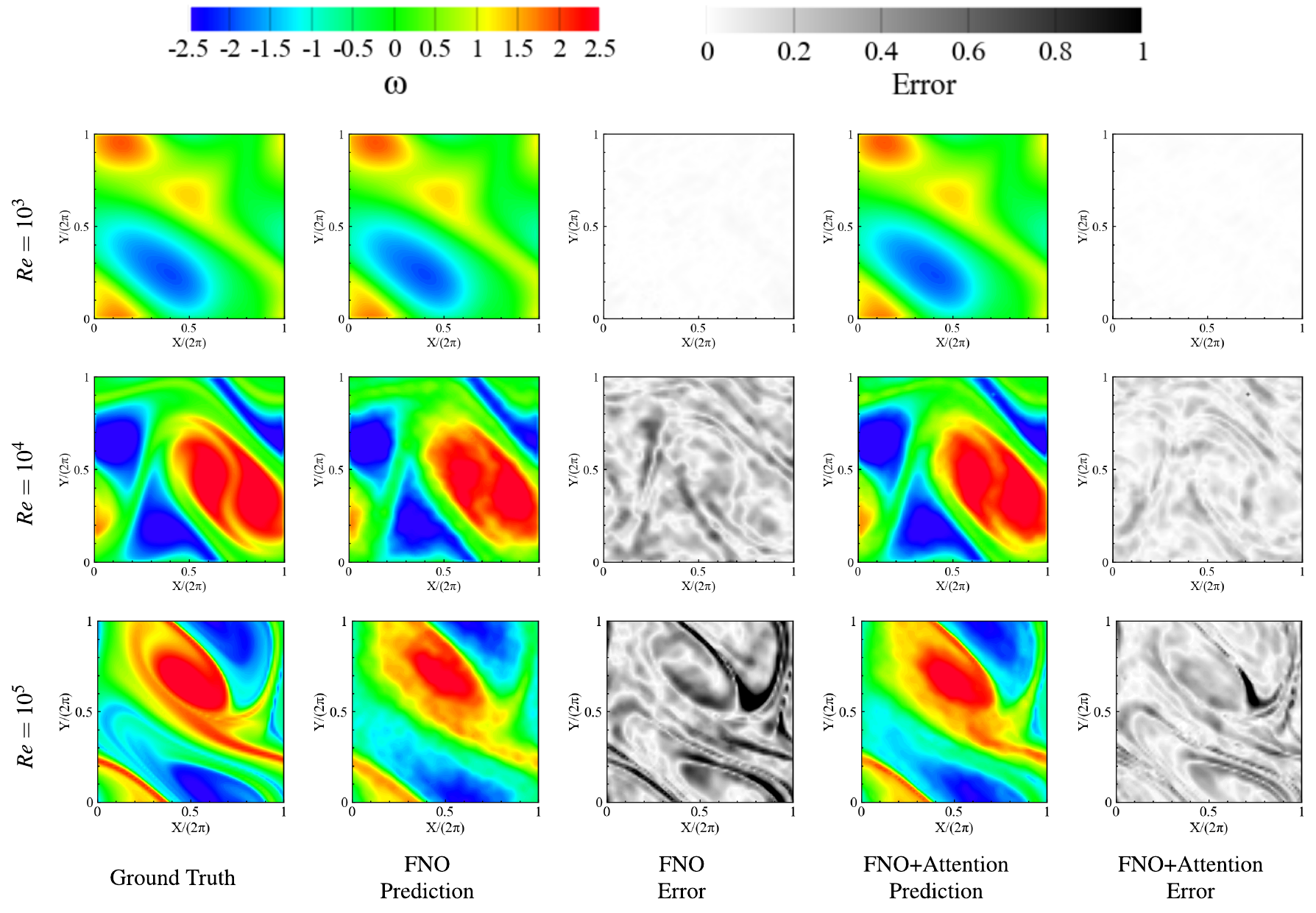}
\caption{Vorticity prediction and absolute error at different Reynolds numbers.}
\label{vor_reno}
\end{figure*}

Fig.\ref{vor_reno} compares the snapshots of predicted vorticity and the absolute errors of the vorticity at different Reynolds numbers. Both models can accurately reconstruct the instantaneous spatial structures of turbulence at small Reynolds number $Re= 10^3$.  However, when the flow becomes more turbulent at higher Reynolds numbers $Re= 10^4$ and $10^5$, the performance improvement of attention-enhanced FNO becomes more significant.

\begin{figure*}
\centering
\includegraphics[width=1\textwidth]{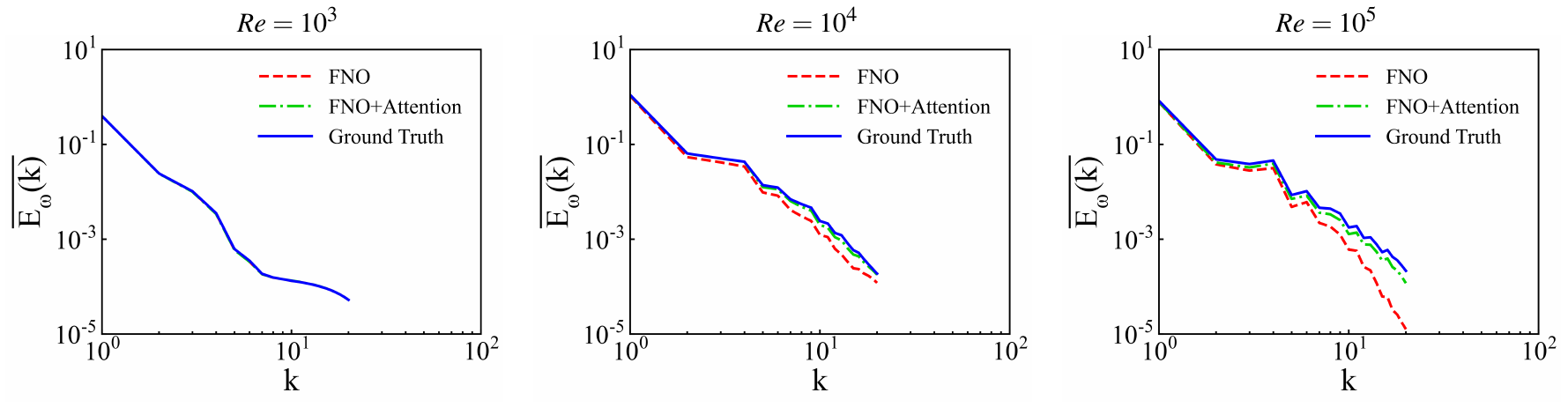}
\caption{Average spectrum at different Reynolds numbers.}
\label{spectra_reno}
\end{figure*}

Fig.\ref{spectra_reno} compares the ensemble-averaged vorticity spectrum of 200 test samples at different Reynolds numbers. At small Reynolds number $Re= 10^3$, both models can reconstruct the multi-scale flow statistics accurately. However, the vorticity spectrum predicted by FNO deviates away from the ground truth as the Reynolds number increases to $Re=10^4$ and $10^5$. In comparison, the attention-enhanced FNO can accurately capture the flow statistics at different scales. 

\begin{figure*}
\centering
\includegraphics[width=1\textwidth]{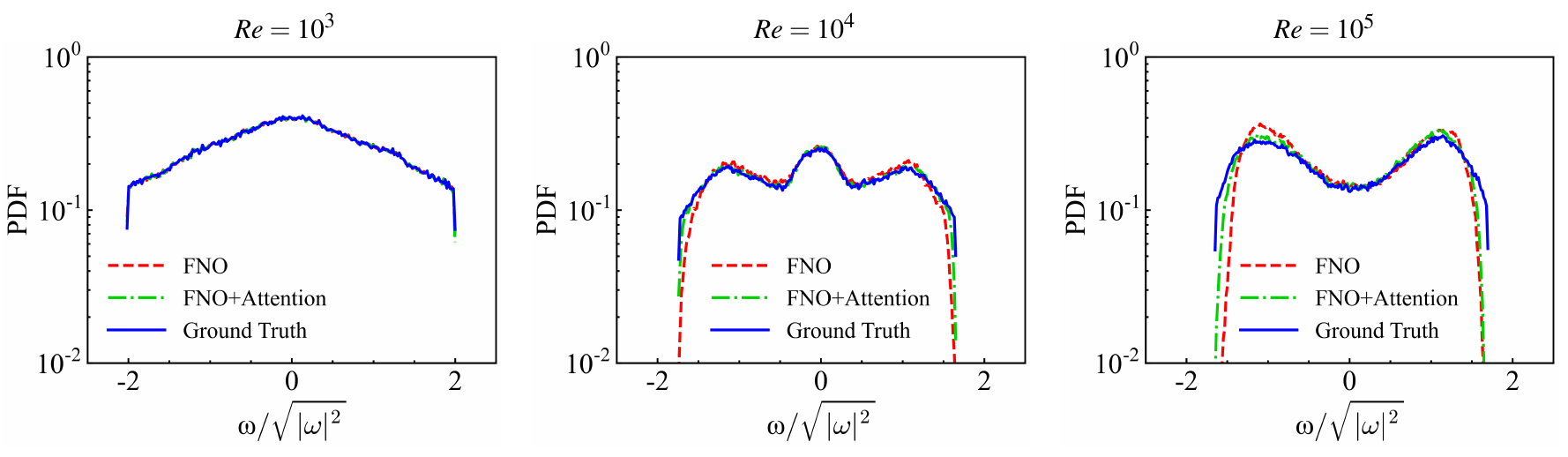}
\caption{PDF of normalized vorticity at different Reynolds numbers.}
\label{pdf_reno}
\end{figure*}

Fig.\ref{pdf_reno} illustrates the PDFs of normalized predicted vorticity at different Reynolds numbers.
At small Reynolds number ($Re = 10^3$), the predictions of both models have good agreements with the ground truth PDF. As the Reynolds number increases to $Re = 10^4$, the predicted PDFs of both models get narrower than the ground truth; and the gap is further enlarged as the Reynolds number increases to $Re = 10^5$.

\begin{figure*}
\centering
\includegraphics[width=1\textwidth]{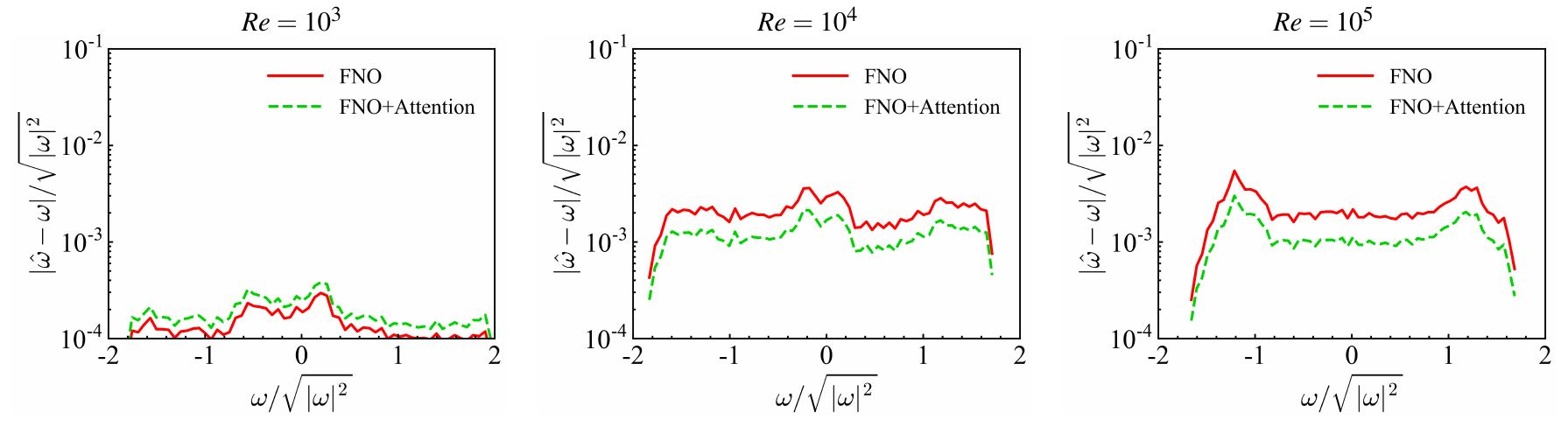}
\caption{Condition average error of vorticity at different Reynolds numbers.}
\label{error_con_avg_reno}
\end{figure*}

Fig.\ref{error_con_avg_reno} shows the conditional average of normalized absolute error at different Reynolds numbers. The prediction errors of both models become larger as the Reynolds number increase from $Re = 10^3$ to $Re = 10^4$ and $Re = 10^5$, and the performance improvement of attention-enhanced FNO also becomes more obvious with the increasing of Reynolds numbers.

\begin{figure}
\centering
\includegraphics[width=0.8\linewidth]{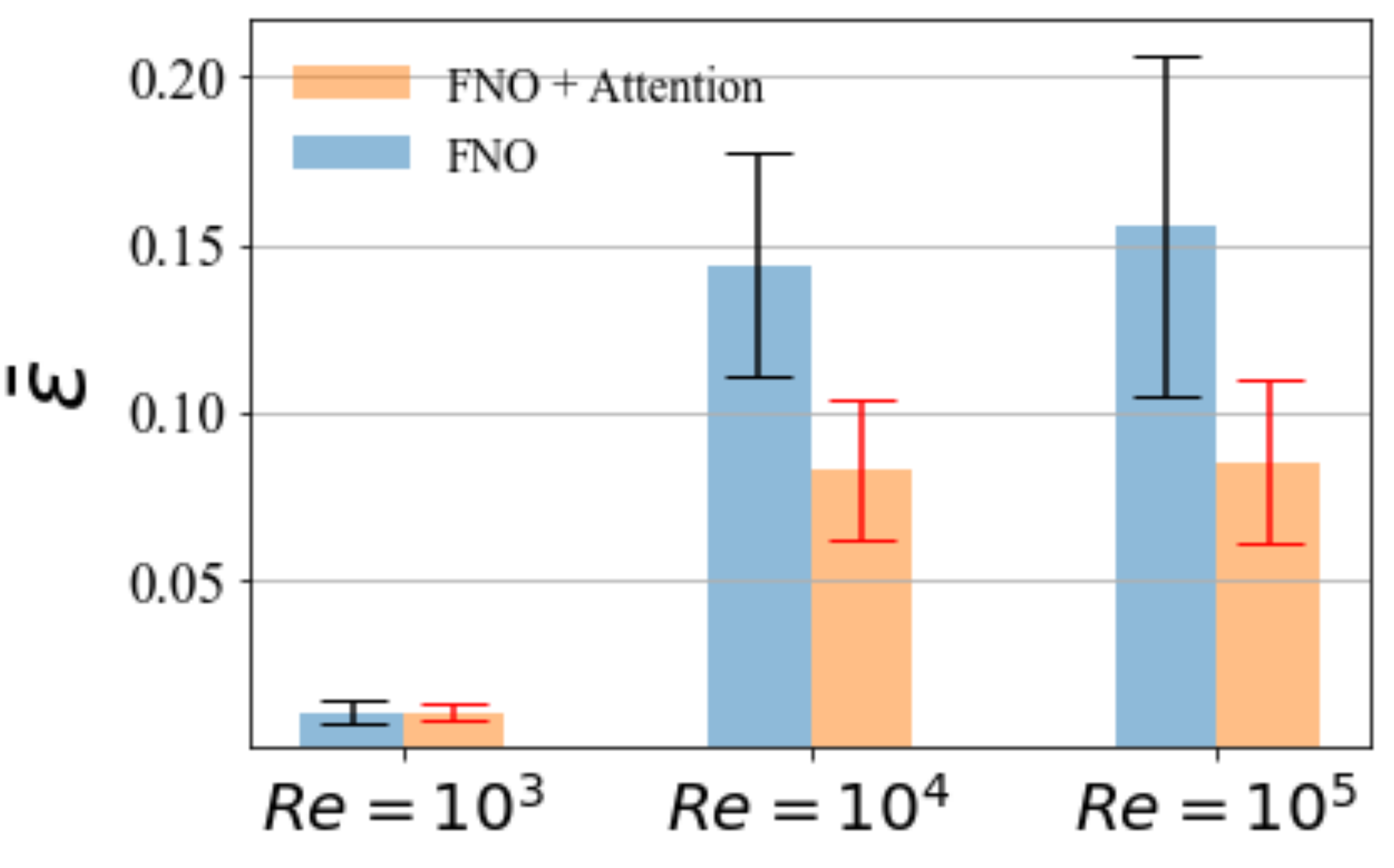}
\caption{Average relative error of vorticity at different Reynolds numbers.}
\label{avg_error_reno}
\end{figure}

Fig.\ref{avg_error_reno} compares the mean and standard deviation of relative error on 200 test samples. Both models can achieve accurate predictions (1\% error) at small Reynolds number  $Re=10^3$; however, as the Reynolds number increases from $10^3$ to $10^4$, the FNO prediction error arises from 1\% to around 15\%, whereas the FNO+Attention prediction error is about 8\%; when the Reynolds number increases from $10^4$ to $10^5$, the mean errors of both models are nearly unchanged except that the error standard deviation becomes larger.

\begin{figure*}
\centering
\includegraphics[width=1\textwidth]{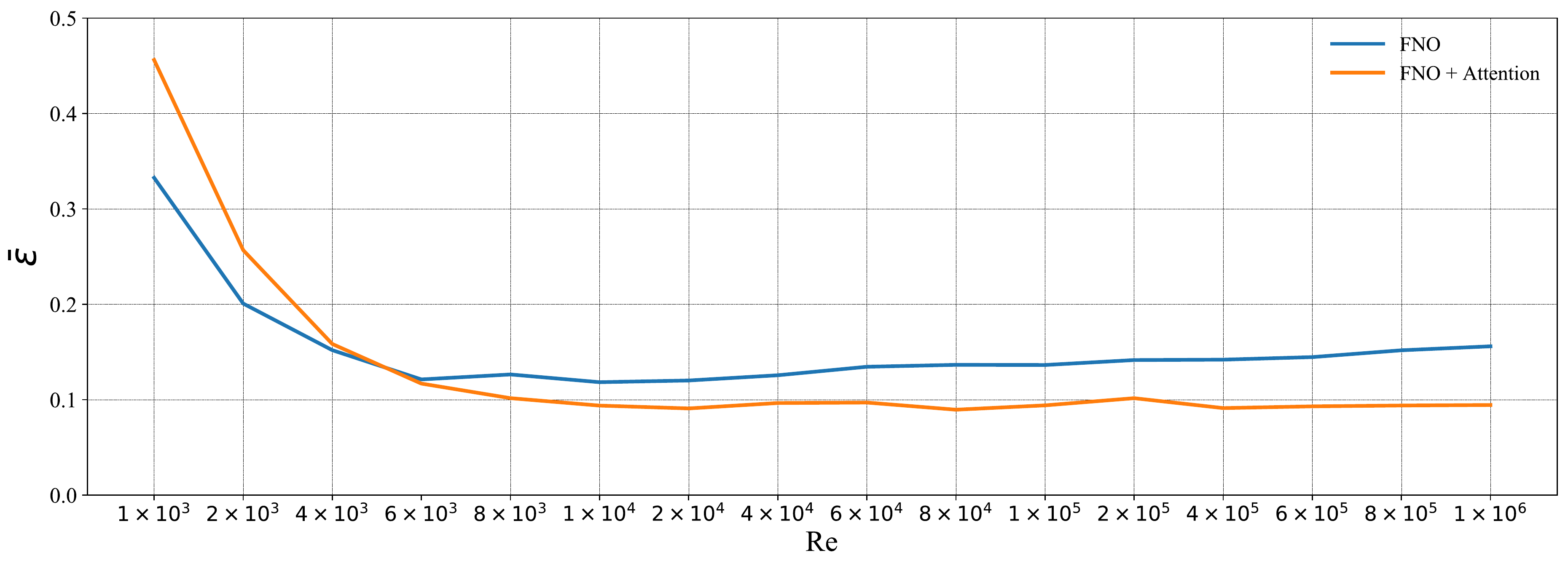}
\caption{Generalization error at different Reynolds numbers ~(models trained on $Re = 1\times10^5$ dataset).}
\label{re_gene2}
\end{figure*}

\begin{figure*}
\centering
\includegraphics[width=1\textwidth]{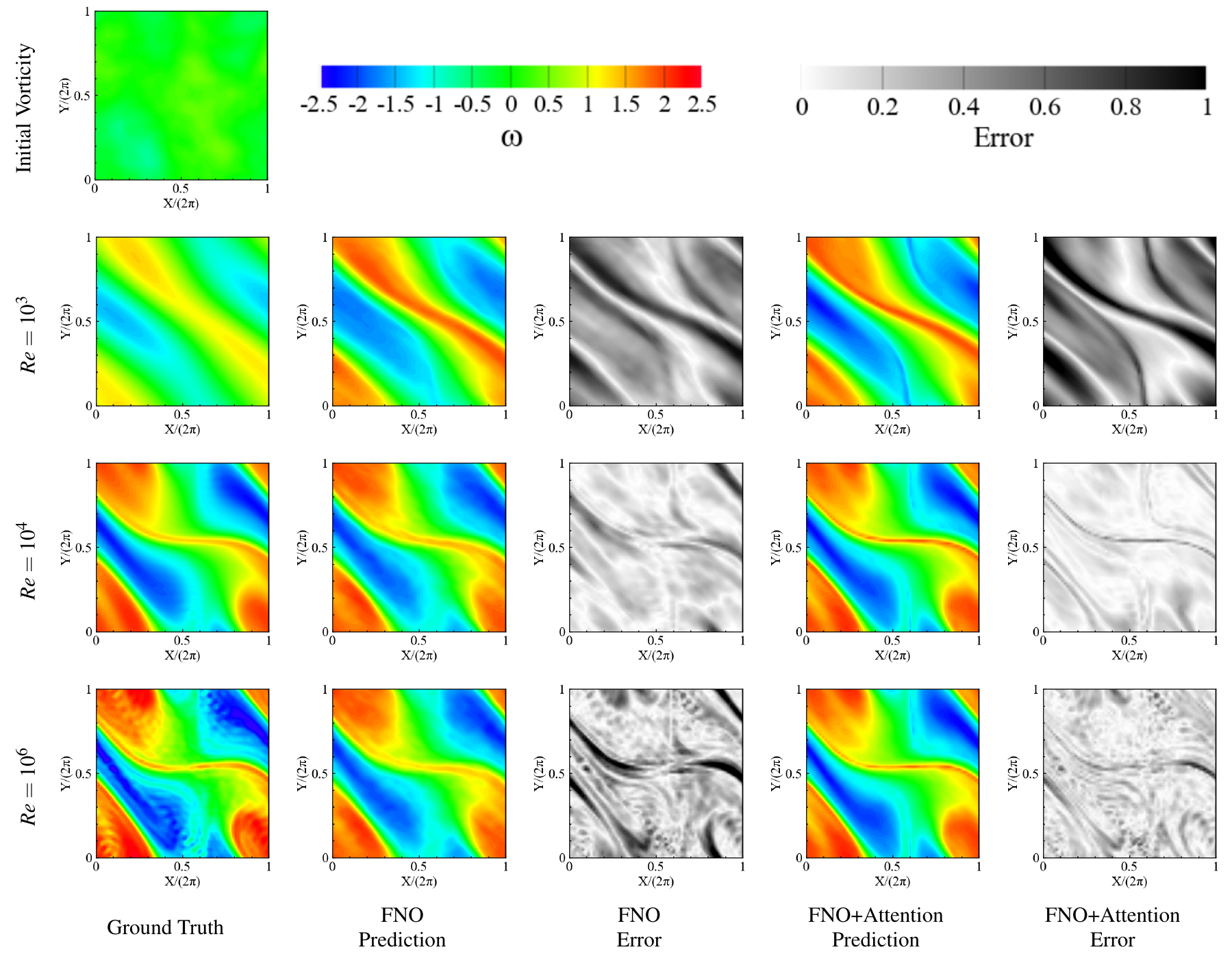}
\caption{Vorticity prediction and absolute error at different Reynolds numbers. ~(models trained on $Re = 1\times10^5$ dataset).}
\label{re_gene3}
\end{figure*}

In addition to benchmark the approximation capacity of the two models, we also benchmark the model generalization ability across different Reynolds numbers. We train both models on $Re = 10^5$ dataset, and test the trained models on unseen Reynolds numbers ranging from $Re = 10^3$ to $Re = 10^6$.\\ \indent Fig.\ref{re_gene2} shows the generalization errors of FNO and attention-enhanced FNO models: both models can generalize well at high Reynolds numbers, whereas FNO generalizes better at low Reynolds numbers (below $Re = 5\times10^3$), and attention-enhanced FNO generalizes better at high Reynolds numbers. This result shows that the attention mechanism can capture the multi-scale characteristics of the high Reynolds numbers turbulence better as we expected. \\ \indent Fig.\ref{re_gene3} shows the vorticity prediction and absolute error where the models are trained on $Re = 1\times10^5$ dataset and tested on three  samples with different Reynolds numbers at $t=15$. The test samples are initialized with the same random state, where the corresponding Reynolds numbers are set to $Re = 10^3$, $Re = 10^4$, and $Re = 10^6$ respectively. Both models fail to reconstruct the flow structures at low Reynolds number $Re = 10^3$, with FNO having smaller error than attention-enhanced FNO. By contrast, when the models are tested at high Reynolds numbers $Re = 10^4$ and $Re = 10^6$, the attention-enhanced FNO produces smaller generalization error than FNO, and can accurately reconstruct the multi-scale flow structures of turbulence.

\subsection{Zero-shot super-resolution benchmark}\label{mesh}

The Fourier layers are discretization-invariant because they can learn from and evaluate functions which are discretized in an arbitrary way. Since parameters are learned directly in Fourier space, resolving the functions in physical space simply amounts to projecting on the basis $e^{2 \pi i\langle x, k\rangle}$ which are well-defined everywhere on $\mathbb{R}^{d}$. This allows zero-shot super-resolution without losing accuracy \cite{li2020fourier}. Such property provides free training acceleration and significant memory saving by training the model on the downsampled data.\\

Since the attention block, as shown in Fig.\ref{attention_block}, adopts $1\times1$ convolution, it retains the FNO mesh-invariant property. In this section, we benchmark the  zero-shot super-resolution performance of two models. We generate data on grid size of $256\times256$ at the Reynolds number of $10^5$, then train both models on the downsampled data of grid size $64\times64$. The trained models are evaluated at the same time step (t=15), on grid resolution of $64\times64$ , $128\times128$ and $256\times256$, respectively. 

\begin{figure*}
\centering
\includegraphics[width=1\textwidth]{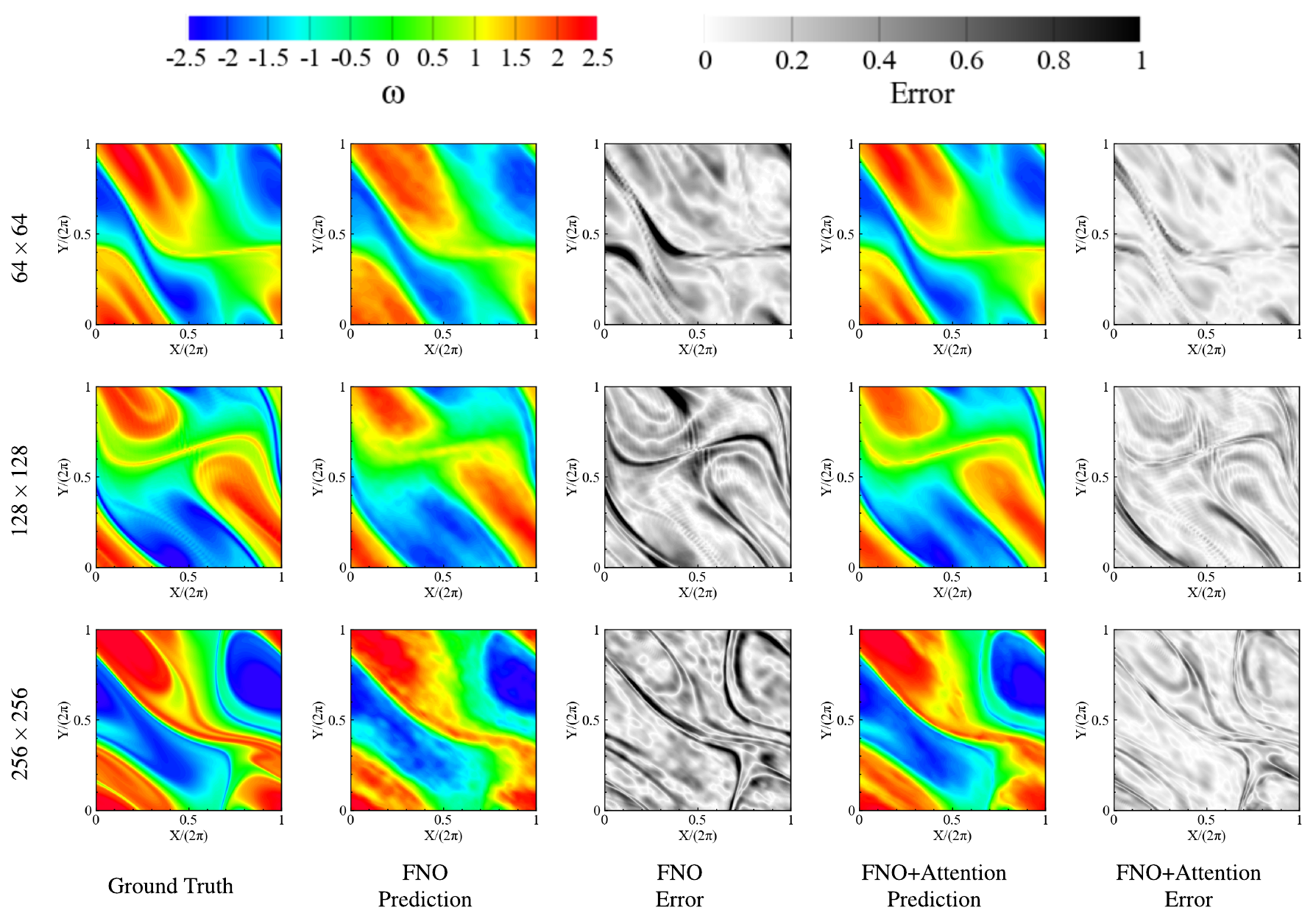}
\caption{Vorticity prediction and absolute error at different resolutions.}
\label{vor_grid}
\end{figure*}

Fig.\ref{vor_grid} shows the snapshots of predicted vorticity and the absolute errors at different grid resolutions. Unlike Figs.\ref{vor_time_steps} and \ref{vor_reno} where the errors increase significantly with time steps and Reynolds numbers, the error rates of both models do not increase with grid size. 

\begin{figure}
\centering
\includegraphics[width=0.8\linewidth]{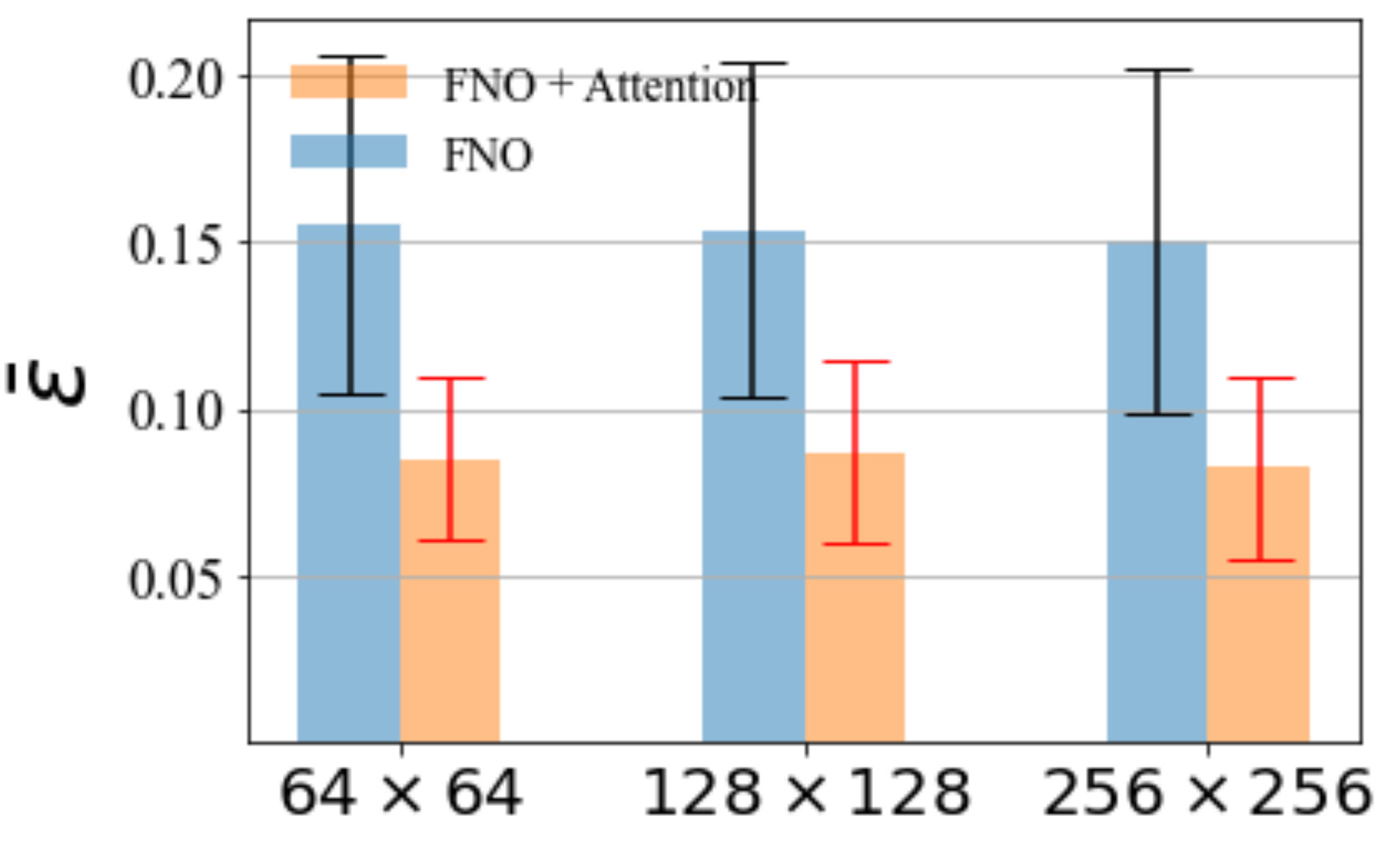}
\caption{Average relative error at different resolutions.}
\label{avg_error_grid}
\end{figure}

Fig.\ref{avg_error_grid} compares the average relative error on different resolutions: both models have consistent error rates among different resolutions, whereas the attention-enhanced FNO achieves over 40\% error reduction at all grid resolutions.

\begin{figure*}
\centering
\includegraphics[width=1\textwidth]{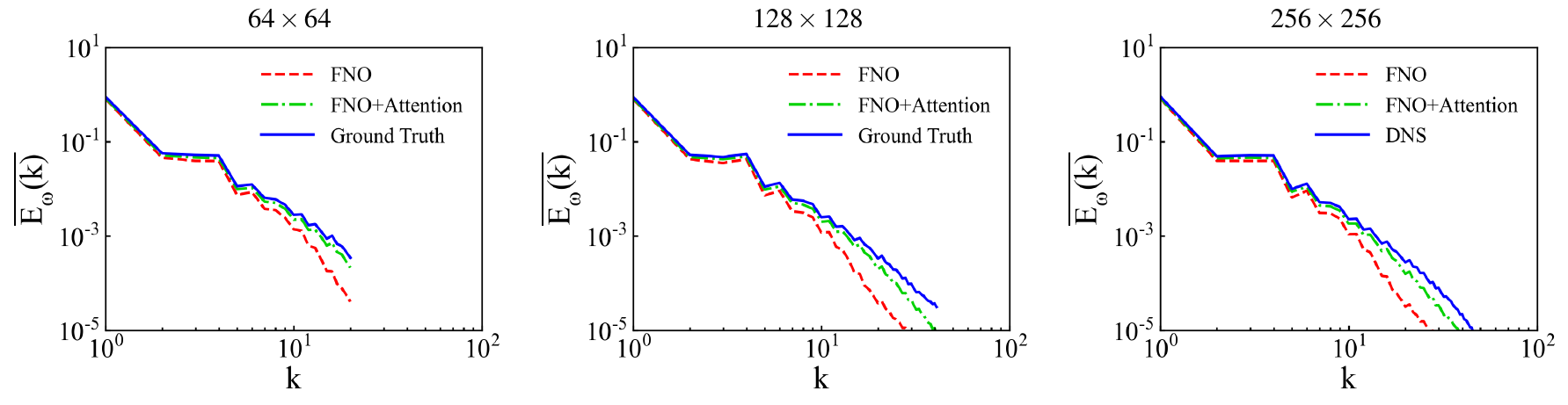}
\caption{Average spectrum of vorticity at different resolutions.}
\label{spec_grid}
\end{figure*}

Fig.\ref{spec_grid} compares the averaged vorticity spectrum of 200 test samples at different grid resolutions: the predicted vorticity spectrum of both models can agree well with the ground truth in low-wavenumber region. However, the FNO predicted spectrum deviates significantly from the ground truth at high-wavenumber region. In contrast, the FNO+Attention model can reconstruct the multi-scale flow structures accurately at all grid resolutions.

\subsection{Performance benchmark on Taylor-Green turbulence and free shear turbulence}\label{turbulence}
In this section, we benchmark the performance of different models on more challenging turbulence simulation tasks: the Taylor-Green turbulence and free shear turbulence. The Taylor-Green and free shear turbulence are governed by the same Navier-Stokes equations displayed in Eq. (\ref{eq:ns}) without the forcing term. The Reynolds number is $Re=10^3$ in both the Taylor-Green turbulence and free shear turbulence. The initial conditions of Taylor-Green and free shear turbulence are given by Eq.\ref{tgv_ic} and Eq.\ref{khi_ic} respectively
\cite{san2018stratified,sengupta2018non}.

\begin{equation}\label{tgv_ic}
\omega \left( {\bm{x},0} \right) = 2\cos 2{x_1}\cos 2{x_2} + \lambda_1,
\end{equation}

\begin{equation}\label{khi_ic}
\omega \left( {\bm{x},0} \right) = 2\lambda_1 \cos 2{x_1} + \delta \left( {\left| {{x_2}} \right|,\frac{\pi }{2}} \right) +\lambda_2.
\end{equation}

Here, the magnitudes of perturbation $ \lambda_1 $ and $ \lambda_2$ satisfy the Gaussian random distribution where $ \lambda_1, \lambda_2 \sim \mathcal{N}\left( {0,{{10}^{ - 3}}} \right)$, and $\delta \left( {a,b} \right)$ is the Kronecker delta function, as described in Eq. \ref{delta}.\begin{equation}\label{delta}
\delta(a, b)= \begin{cases}1 & \text { if } a=b \\ 0 & \text { otherwise }\end{cases}
\end{equation}\\\indent Since it takes a long time for the Taylor-Green vortex to transition from the laminar flow to turbulence, we rotate the initial Taylor-Green vortex to a certain angle ($\theta  \in \left( { - {{10}^ \circ },{{10}^ \circ }} \right)$) to break the symmetry of the base flow\cite{san2018stratified}. We generate 1200 data samples on the grid size of $64\times64$, where each sample contains 30 steps of solutions of a random initialized condition. Both models are trained on $t=0$ to $t=20$, and are tested on $t=11$ to $t=30$ to benchmark generalization performance on time dimension.

% The data generation procedure and gird size are consistent with those of the 2D incompressible isotropic turbulence. 

\begin{figure*}
\centering
\includegraphics[width=1\textwidth]{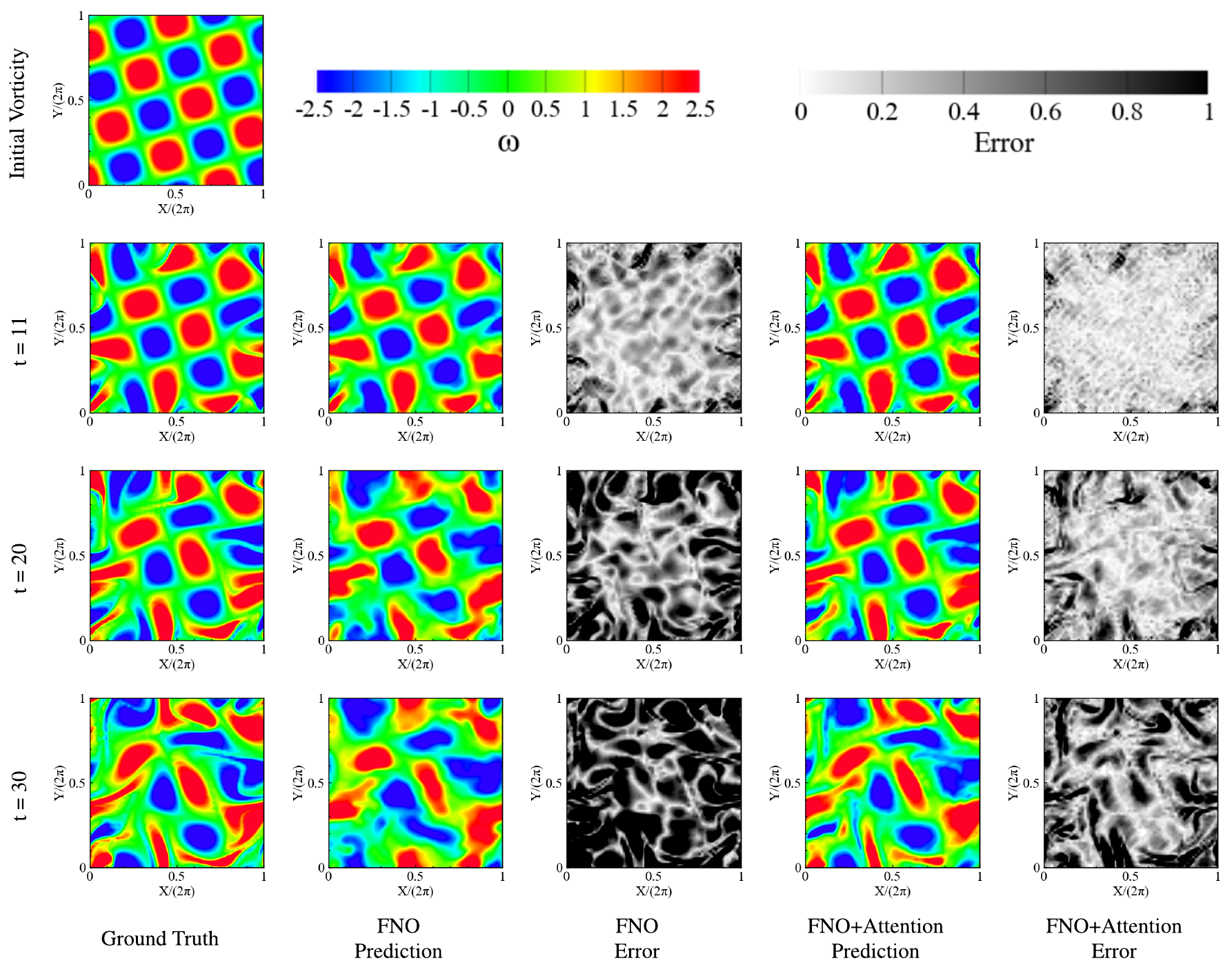}
\caption{Vorticity prediction and absolute error at selected time steps~(Taylor-Green turbulence).}
\label{Taylor-Green_vor}
\end{figure*}

Fig. \ref{Taylor-Green_vor} compares the predicted vorticity and the absolute errors on a test sample of Taylor-Green turbulence. The prediction errors of both models increase with the advance of time. At $t=11$, both models can accurately reconstruct the instantaneous spatial structures of turbulence in the beginning; at $t=20$, the attention-enhanced FNO can still make accurate reconstructions whereas the FNO cannot; at $t=30$ where both models are tested on unseen time steps, the errors become too large for both models to perform accurate reconstructions.

\begin{figure*}
\centering
\includegraphics[width=1\textwidth]{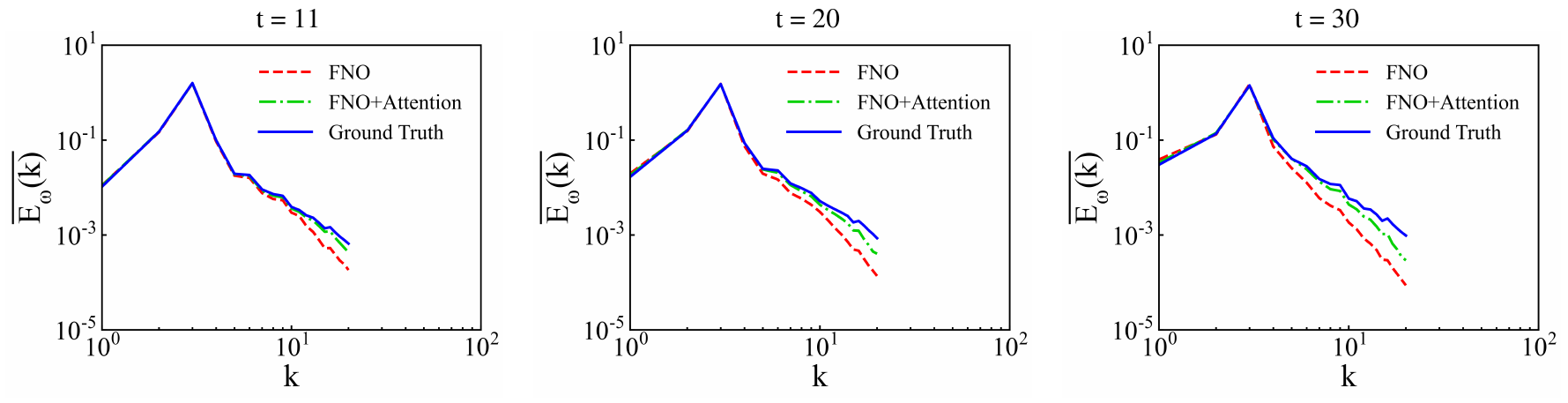}
\caption{Averaged vorticity spectrum on 200 test samples (Taylor-Green turbulence).}
\label{tgv_spectra_time_steps_avg}
\end{figure*}

Fig.\ref{tgv_spectra_time_steps_avg} shows the ensemble-averaged vorticity spectrum $\overline{E_{\omega}(k)}$ of Taylor-Green flow on 200 test samples. At $t=11$, the predicted vorticity spectrum of FNO can agree well with the ground truth at low wave numbers, but deviates from the ground truth at high wave numbers; whereas the attention-enhanced FNO can well reconstruct the vorticity spectrum at different flow scales by accurately capturing small-scale flow structures. As time advances to $t=20$, both models deviate further away from the ground truth at high-wave number region, with attention-enhanced FNO having better agreement with the ground truth. At $t=30$, the predicted vorticity spectrum of FNO 
starts to deviate from the ground truth at low-wave number region too, whereas the predicted vorticity spectrum of attention-enhanced FNO does not.

\begin{figure*}
\centering
\includegraphics[width=1\textwidth]{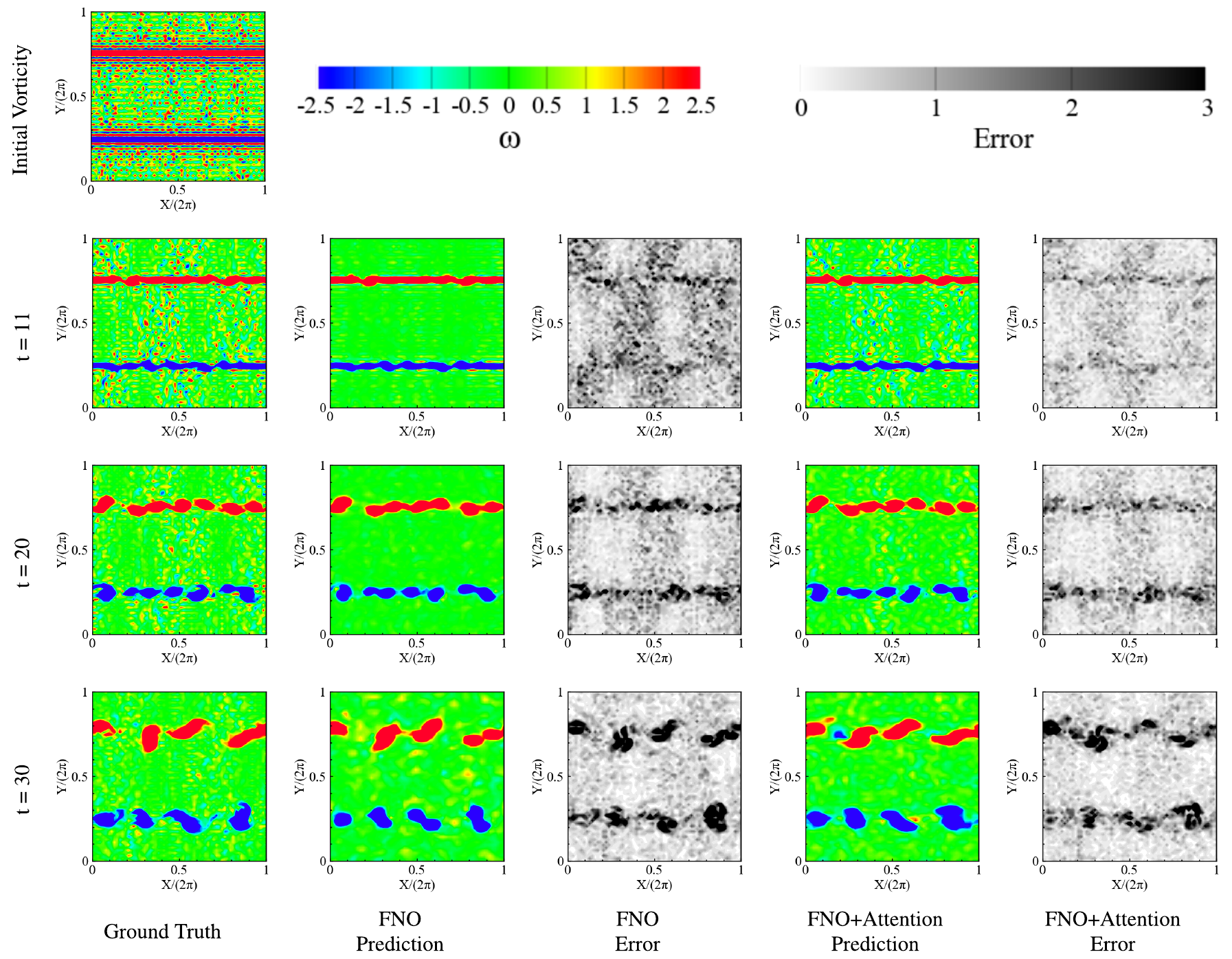}
\caption{Vorticity prediction and absolute error at selected time steps~(free shear turbulence).}
\label{free_shear_vor}
\end{figure*}

\begin{figure*}
\centering
\includegraphics[width=1\textwidth]{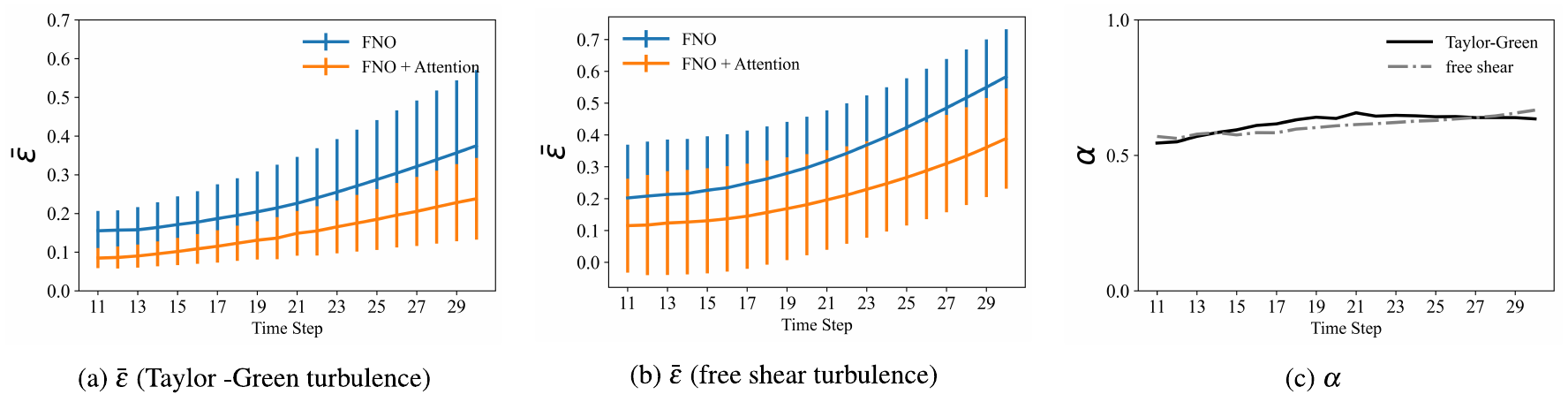}
\textbf{(a)} Spatial-averaged relative error of vorticity on 200 test samples (Taylor-Green turbulence). \textbf{(b)} Spatial-averaged relative error of vorticity on 200 test samples (free shear turbulence). \textbf{(c)} Ratios of mean relative error.
\caption{Relative error comparison at consequent time steps.}
\label{error_time_steps3}
\end{figure*}

Fig.\ref{free_shear_vor} compares the predicted vorticity and the absolute errors on a test sample of the free shear turbulence. The prediction errors of both models increase with the advance of time, with the errors of attention-enhanced FNO being significantly smaller than FNO in terms of the region and magnitude. At $t=11$ and $t=20$, the attention-enhanced FNO captures the small-scale structures better than FNO, whereas both of them fail to reconstruct the instantaneous large-scale structures at $t=30$.\\ \indent Fig.\ref{error_time_steps3} presents the spatial-averaged relative errors of two models on predictions of the Taylor-Green turbulence and free shear turbulence. For the Taylor-Green turbulence, both models can generalize well on unseen time steps from $t=21$ to $t=30$; whereas for the free shear turbulence, the prediction errors  increased significantly from $t=21$ to $t=30$, indicating that both models fail to generalize well on time dimension. In both Taylor-Green and free shear turbulence, the attention-enhanced FNO achieves 40\% error reduction throughout all time steps.

\section{Discussion}\label{discussion}
\subsection{Why does attention work ?}
One of the most common criticisms that NNs often face is the lack of interpretability, therefore they are often treated as ``black-box'' surrogate models. Understanding how neural network based models make inference on physical problems requires future efforts from both data scientists and specialized experts of the area. We present some thoughts and ideas from the perspective of turbulence modeling, and leave them as an open discussion.

\begin{figure*}
\centering
\includegraphics[width=1\textwidth]{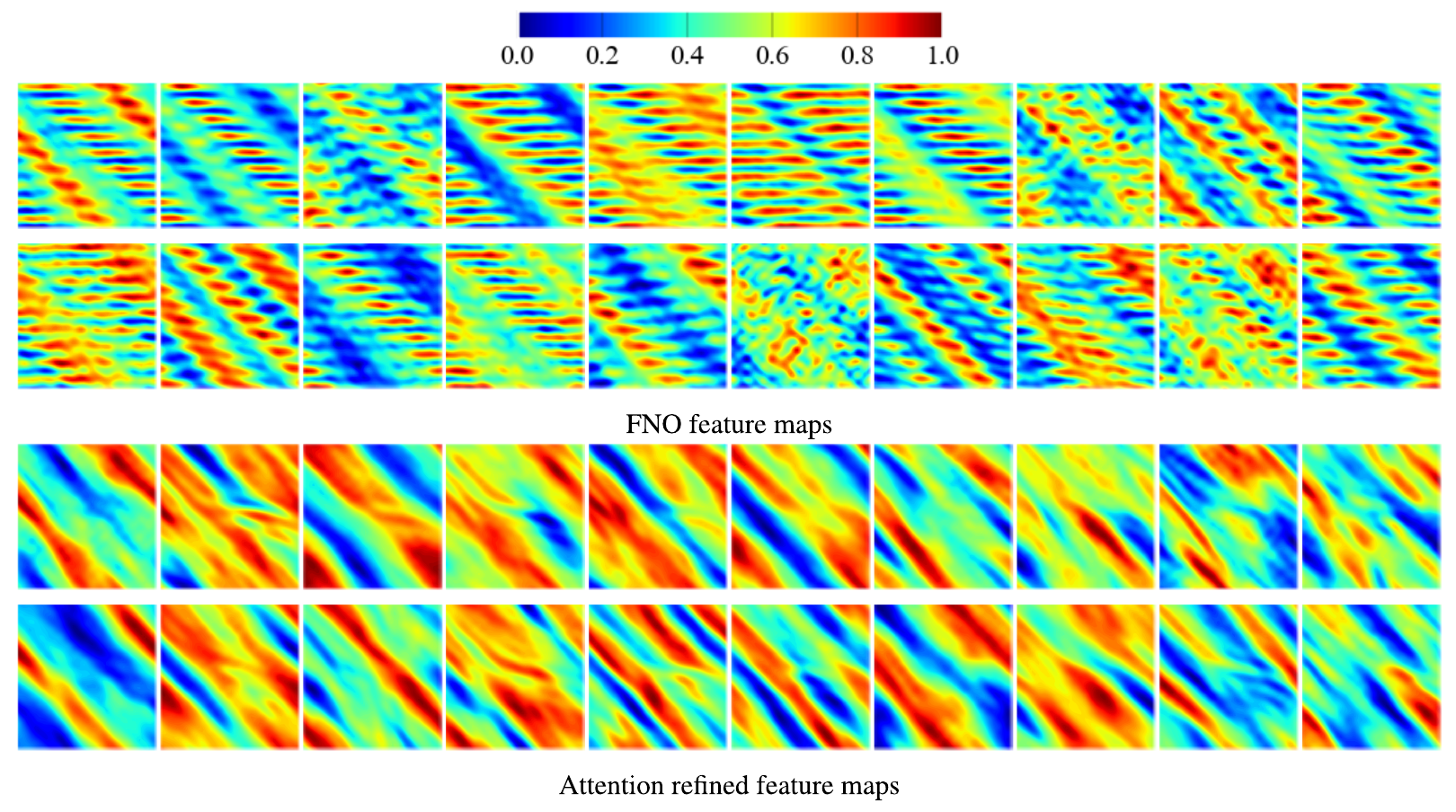}
\caption{Comparison of feature maps.}
\label{feature maps}
\end{figure*}

\begin{figure}
\centering
\includegraphics[width=0.8\linewidth]{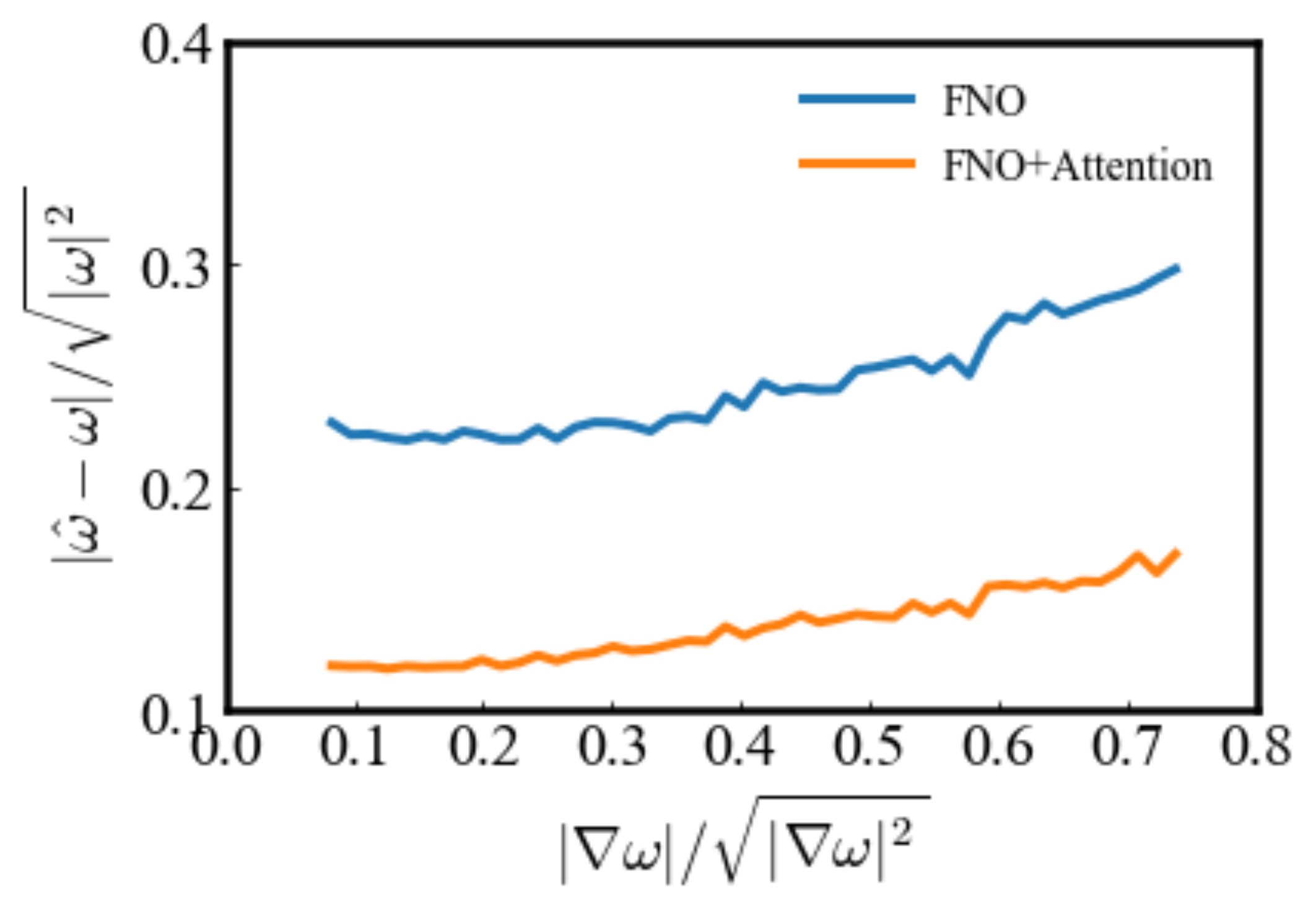}
\caption{Normalized absolute error versus normalized
vorticity gradient.}
\label{error_grad}
\end{figure}

\begin{itemize}
    \item We have investigated the FNO feature maps $v_{t}(\textbf{x})$ and the attention refined feature maps $v_{t}(\textbf{x})'$. These feature maps are the output of an intermediate layer, and they contain the information of how the neural networks understand turbulence. The feature maps are stacked together and are characterized by a 3-dimensional metrics ($C\times S\times S$), where $C$ is the channel dimension and $S$ is the grid size. The value of the channel dimension equals to the number of convolution filters used in the Fourier layer (20 in this work). Fig.\ref{feature maps} shows the unfolded feature maps of each channels. We noticed that the features learned by FNO are relatively uniformly distributed in space, whereas the attention refined features are not. Moreover, the attention refined feature maps are visually similar to the vorticity distribution, indicating that attention module can capture the turbulence nonequilibrium features better as we expected.
    
    \item Another interesting phenomenon is that the large error is closely related with the regions where the vorticity changes dramatically, as shown in Fig.\ref{vor_time_steps}. We therefore  investigate the relationship between the absolute error $|\hat{\omega}-\omega|$ and vorticity gradient $\nabla\omega$, as shown in Fig.\ref{error_grad}. We noticed that the errors of both models grow with the increasing of vorticity gradient. Further improvements can be made by adding an extra vorticity gradient penalty term with the loss function during training. 
\end{itemize}

\subsection{Computational efficiency}
Table \ref{tab:benchmark} compares computational cost of 10 prediction steps on a $64\times64$ grid using 3 different approaches. These numerical experiments are ran on a virtual machine powered by Google Colab, where the CPU type is Intel(R) Xeon(R) CPU @ 2.30GHz and GPU type is Tesla K80. Once trained, the surrogate models can be extremely efficient compared with traditional approaches. Both models provide 8000 folds speedup compared with the pseudo-spectral numerical solver.  The attention-augmented neural network model achieves 40\% error reduction at the same level of memory consuming and computational expense.

\begin{table}
    \begin{center}
    \def~{\hphantom{0}}
\begin{tabular}{l|c|c} 
Method & Parameters  & GPU Timing\\
\hline
Numerical Solver & N/A  & $502.85 s$  \\ 
FNO & 465,717  & $0.0579 s$  \\
FNO + Attention & 466,222 & $0.0587 s$  \\ 
\end{tabular}
    \caption{Computational efficiency comparison of three different numerical approaches.}
    \label{tab:benchmark}
    \end{center}
\end{table}

\section{Conclusion}\label{conclusion}

In this work, we propose an attention-enhanced neural network approach to model the nonequilibrium feature of turbulence. Numerical experiments show that: 1) the proposed model can significantly reduce the prediction error induced by temporal accumulation, and can accurately reconstruct the instantaneous spatial structures of turbulence.  2) the attention-enhanced FNO model can reduce the prediction error rate by 40\% as the flow becomes more turbulent at higher Reynolds numbers. 3) the attention-enhanced model retains the mesh-invariant property of FNO: the FNO+Attention model can be trained on lower grid resolution and evaluated on higher grid resolution, without losing accuracy. 4) The attention enhanced FNO model achieves the same level of computational efficiency as compared with the original FNO model.  

\section{Data AVAILABILITY}
The data that support the findings of this study are available
from the corresponding author upon reasonable request.
\section*{Acknowledgments}
This work was supported by the National Natural Science Foundation of China (NSFC Grant Nos. 91952104, 92052301, 12172161 and
91752201), by the National
Numerical Windtunnel Project (No.NNW2019ZT1-A04), by the Shenzhen Science and Technology Program (Grants No.KQTD20180411143441009), by Key Special Project for Introduced Talents Team of Southern Marine Science and Engineering Guangdong Laboratory (Guangzhou) (Grant No. GML2019ZD0103), by CAAI-Huawei MindSpore Open Fund, and by Department of Science
and Technology of Guangdong Province (No.2020B1212030001). This work was also
supported by Center for Computational Science and Engineering of Southern University of Science and Technology.

\bibliography{aipsamp}% Produces the bibliography via BibTeX.

\end{document}